\documentclass[12pt]{iopart}

\usepackage{iopams}
\usepackage{graphicx}%
\usepackage[english]{babel}
\usepackage{bm}
\usepackage{graphicx}
\usepackage{braket}
\usepackage{xcolor}
\usepackage{booktabs}
\usepackage{letltxmacro}
\usepackage[normalem]{ulem}
\usepackage{longtable}
\usepackage{dcolumn}
\usepackage{multirow}

\definecolor{orange}{rgb}{1,0.5,0}

\begin{document}

\title{Helmholtz Fermi Surface Harmonics: an efficient approach for treating anisotropic problems involving Fermi surface integrals}

\author{Asier Eiguren$^{1,2}$ and Idoia G Gurtubay$^{1,2}$}
\address{$^1$Materia Kondentsatuaren Fisika Saila, Zientzia eta Teknologia Fakultatea,
Euskal Herriko Unibertsitatea UPV/EHU, 644 Postakutxatila, E-48080 Bilbo, Basque Country, Spain}
\address{$^2$Donostia International Physics Center (DIPC), Paseo de Manuel Lardizabal,
                                             E-20018, Donostia, Basque Country, Spain}

\date{\today}

\begin{abstract}
We present a new efficient numerical approach for representing anisotropic 
physical quantities and/or matrix elements defined on the Fermi surface of metallic materials.
The method introduces a set of numerically calculated generalized orthonormal functions 
which are the solutions of the Helmholtz equation defined on the Fermi surface.
Noteworthy, many properties of our proposed basis set are also shared by the 
Fermi Surface Harmonics (FSH) introduced by Philip B. Allen 
[Physical Review B $\mathbf{13}$, 1416 (1976)], proposed to be constructed as polynomials
of the cartesian components of the electronic velocity.
The main motivation of both approaches is identical, to handle anisotropic problems efficiently.
However, in our approach the basis set is defined as the eigenfunctions of a differential operator and several
desirable properties are introduced by construction. 
The method demonstrates very robust in handling problems with any crystal structure or topology of the Fermi surface, and the periodicity of the reciprocal
space is treated as a boundary condition for our Helmholtz equation. 
We illustrate the method by analysing the free-electron-like Lithium (Li), Sodium (Na), Copper (Cu),
Lead(Pb), Tungsten (W) and Magnesium diboride (MgB$_2$). 
\end{abstract} 


\date{\today}

\maketitle

\section{Introduction}

The Fermi surface (FS) of a metal is a characteristic property of the crystal structure and the material itself.
It is found in many physical problems that the exact shape, topology and the precise value of a scalar,
 vector or 
tensor variable defined on this surface is crucial for a good description of any physical model. 
The importance of an accurate and efficient method for storing, filtering or interpolating a scalar or vector physical 
quantity on a Fermi surface is demonstrated very clearly in the prototype examples of transport and in the electron-phonon (EP) problem.
In the EP interaction, 
the vibrational frequencies are typically about one/two orders of magnitude smaller than the electron energies, 
and to a good approximation, the interaction is described by scattering 
events connecting different points of the Fermi surface  $\mathbf{k}$ and $\mathbf{k'}$. The EP
 theory requires the knowledge of a huge amount of electron-phonon matrix elements 
$g^{\nu}_{\mathbf{k},\mathbf{k'}}$ connecting different points of the Fermi surface mediated by several phonon branches ($\nu$).
For example, one needs typically about $n_k$$\sim$10$^4$ data points for a good resolution of the matrix elements defined on the Fermi surface. 
In a rough estimation for a simple EP problem, considering  $n_p$$\sim$$10$ phonon branches and about $n_e$$\sim$$10$ electron bands, 
the  required amount of data for the matrix elements is then $N$ $\sim$ 10$^{10}$. 
Thus, filtering or compressing all this data while keeping accuracy appears very appealing. 

The essence of the present method is to perform a linear integral transformation on quantities such as
 the EP matrix elements
$g^{\nu}_{\mathbf{k},\mathbf{k'}}$ $\rightarrow$ $\tilde{g}^{\nu}_{L,L'}$ 
so that the dimension of the new basis set is smaller than the original one ($n_L$$\sim$10$^2$ 
instead of $n_k$$\sim$10$^4$).

The idea of representing scalar quantities as a function of an orthogonal set defined on the Fermi surface
was introduced by Philip B. Allen in 1976  with the so-called Fermi Surface harmonics (FSH) \cite{FSH-Allen,AllenMitrovic}.
Allen considered a functional set constructed as polynomials of the 
cartesian components of the electronic velocity, and designed -in principle- to operate with any crystal structure and number of Fermi sheets.
The author was able to rewrite the electron self energy, transport equations and the Eliashberg theory of the superconductivity in 
terms of the FSH set. Although the great potential of the FSH method by Allen appeared promising at first, 
the above method has not found a systematic application yet, being applied only in relatively simple systems \cite{Seebeck-Xu}.
The weak impact of the above method might
be  attributed to the several semi-analytic steps involved, 
the relatively complex treatment of different Fermi sheets,  the difficulty to generate the functional set and 
to the fact that the completeness of the latter was not guaranteed. 

We propose a new functional set with a similar spirit and motivation as in Allen's
 FSH \cite{FSH-Allen,AllenMitrovic} but defined very differently,
 and constructed in such a way that the connection with ordinary Fourier transform in flat space and/or with ordinary 
spherical harmonics functions in a (curved) sphere is direct.
Our proposed functional basis is defined to satisfy a modified version of the
Helmholtz equation defined on the Fermi surface. More graphically, in our method the Fermi 
surface is considered as if it were a vibrating membrane. The standing waves 
 calculated on this surface
constitute the new proposed functional set.
As these functions are the solution of a second order partial differential equation, the -generalized- orthogonal property of the
basis set is recovered by construction, and more importantly, the completeness of the set is automatically guaranteed.
Another advantage of the present method, in comparison with the Allen FSH set, is 
that a definition of an energy cutoff $E_c$ for the basis set appears naturally. 
As in ordinary plane wave theory, the plane wave cutoff ($E_c$) allows an estimation of the 
minimal size which is describable by the new basis set. 
Thus, the sharper the target details defined on the Fermi surface, the larger the cutoff and the number of modes 
we need for  its accurate description.
Furthermore, the global and topological properties  of the surface are included automatically  
as demonstrated, for example, by considering a direct application of the Gauss-Bonnet theorem which we utilize as a check.

\section{Mathematical definition and implementation of Helmholtz Fermi Surface harmonics (HFSH)}
\label{sec:definition}

We define de {\it Bare} Helmholtz Fermi Surface harmonics functional basis (BHFSH), as the 
eigenfunction set of the Laplace-Beltrami operator defined on the Fermi surface and
with crystal periodic boundary conditions
\begin{eqnarray} \label{eq:bhelmholtz}
 \nabla_{\mathbf{k}}^2 \Psi_L(\mathbf{k}) + \kappa^2_L \Psi_L(\mathbf{k})=0.
\end{eqnarray}
We stress that $\nabla_{\mathbf{k}}^2$ operates on a 2D surface, 
not in the embedding 3D $\mathbf{k}$-space, and it must be interpreted as in the ordinary Laplace operator, i.e., as the  divergence of gradient.
From now on, we will refer to the Laplace-Beltrami operator simply as the Laplace operator.
In (\ref{eq:bhelmholtz}) $\kappa^2_L$ are the eigenvalues associated to the BHFSH, which 
satisfy the following orthogonality relation:
\begin{eqnarray}\label{ortBHFSH}
\int d^2s_{\mathbf{k}} \Psi_{L'} (\mathbf{k}) \Psi_{L} (\mathbf{k}) = \delta_{L',L} \int d^2s_{\mathbf{k}}.
\end{eqnarray}

In many mathematical problems such as when the unique objective is to 
simply compress a large numerical data set defined on the Fermi surface, 
the set  $\{\Psi_{L} (\mathbf{k}) \}$ might result convenient. 
However, in many physical theories, 
the integrals over the Fermi surface  contain a  weighting factor which is proportional to the local
density of sates, i.e. the inverse of the electron velocity. 
It may be thus desirable to introduce a generalized 
orthogonality condition incorporating the inverse of the electron velocity as a weighting factor. 
Therefore, we define the  Helmholtz Fermi Surface harmonics (HFSH), as the eigenmodes
of a modified version of (\ref{eq:bhelmholtz}), where we introduce the
absolute value of the electron velocity in a given point $\mathbf{k}$ 
of the Fermi surface, $v(\mathbf{k})$:
\begin{eqnarray} \label{eq:helmholtz}
  v(\mathbf{k}) \nabla_{\mathbf{k}}^2 \Phi_L(\mathbf{k}) + \omega_L \Phi_L(\mathbf{k})  =0,
\end{eqnarray}
where $\omega_L$ are the eigenvalues associated to the HFSH set $\{\Phi_L(\mathbf{k}\}$.

As the HFSH elements are the solutions of a second order -linear- partial differential equation, the generalized orthogonality property is now 
\begin{eqnarray}\label{eq:ortHFSH}
\int \frac{d^2s_{\mathbf{k}}}{v(\mathbf{k})} \Phi_{L'} (\mathbf{k}) \Phi_L (\mathbf{k})
= \delta_{L',L} \int \frac{d^2s_{\mathbf{k}}}{v(\mathbf{k})},
\end{eqnarray}
which includes naturally the local density of states as a weighting factor.

Clearly, the simple plane wave basis set ($e^{i \mathbf{k} \cdot \mathbf{r} }$)
and the ordinary spherical harmonics basis set [$Y^m_l(\hat{k})$] are automatically recovered in the HFSH set, defined as above, 
because these functions satisfy the Helmholtz equation in their respective spaces, the flat space and the curved unit sphere, respectively.
Thus, our approach applied in an approximately spherical Fermi surface
with nearly constant density of states leads naturally to the ordinary spherical harmonics. 

The Fermi surface lives in the electron momentum space which is periodic, thus in our
 problem (\ref{eq:helmholtz})
must be solved with the appropriate periodic boundary conditions. 
We will see that in the simple case 
of a single Fermi sheet with no intersection with the boundaries of the first Brillouin zone
 the periodic boundary conditions are automatically satisfied. 
However, for surfaces or multiple Fermi sheets
 intersecting the Brillouin zone boundaries,
 a numerically more elaborated procedure is needed.

\section{Numerical scheme}

In this section we describe the numerical algorithm to solve  (\ref{eq:helmholtz}),
or alternatively  (\ref{eq:bhelmholtz}).

A brief description of the procedure could be the following.
In a first step, a triangulated mesh of the Fermi surface is constructed, such that the  
discretized version of our generalized Helmholtz equation given by (\ref{eq:helmholtz})
 can be solved numerically.  
Once the triangular tesselation of the surface is obtained, a discrete version of the Laplace 
operator can be derived, as described for instance in \cite{LB-Mayer,LB-Ch}, 
and in this way, the original partial differential equation is transformed into  
a generalized sparse eigenvalue problem. The solutions of this linear problem 
are exactly our proposed HFSH set.

\subsection{Discretization of the problem: Triangular tesselation of the Fermi surface} 

As already mentioned, the construction of the triangular mesh over the Fermi 
surface allows us to define an approximate discrete version of the Helmholtz equation and represents 
a very important step when defining all the variables involved in the computation, i.e., the triangle areas,
internal angles, local curvature, etc.     

For this purpose, we have implemented both the marching cube and the marching tetrahedra algorithms \cite{Visualization} (see  \ref{marchingt}). 
The marching cube algorithm is more popular and generates a smaller amount of triangles compared to the marching tetrahedra, 
but an important drawback of this method is that it includes non-manifold features (holes).
We have found that although the marching tetrahedra method introduces a larger amount of triangle
simplexes, the method shows to be much more robust.

\subsection{Discrete version of the Helmholtz equation in a triangulated surface} 

\begin{figure}
\centerline{\includegraphics[width=0.6\columnwidth]{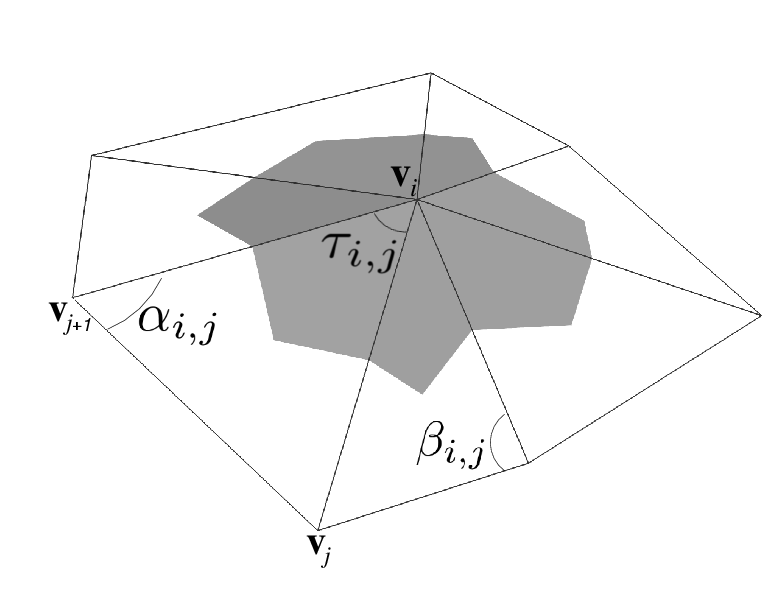}}
\caption{
Schematic representation of the triangulation of the Fermi surface.
Each vertex, $\mathbf{v}_i$, is a 3D $\mathbf{k}$-space vector lying on the Fermi surface
and is surrounded by its nearest neighbour vertices 
$\mathbf{v}_j$. The shaded area represents the barycenter control area around vertex $\mathbf{v}_i$.
 The angles entering the discretized version of the Laplace operator 
in (\ref{eq:disc-lap}), $\alpha_{i,j}$ and  $\beta_{i,j}$, are the opposite angles of the 
 the edge joining vertices $\mathbf{v}_i$ and $\mathbf{v}_j$  shared by   two adjacent triangles.}
\label{fig:baric}
\end{figure}

The triangulation of the Fermi surface -with any method- produces a set of $n_t$ triangles 
whose vertices $\mathbf{v}_i$ ($i$=1,\ldots, $n_v$) are  3D $\mathbf{k}$-space vectors obtained
 by the triangulation process and $n_v$ are the number of vertices. 
Figure \ref{fig:baric}  illustrates the situation. Each vertex ($\mathbf{v}_i$) has  $N_n(\mathbf{v}_i)$ 
nearest neighbour vertices ($\mathbf{v}_j$) and nearest neighbour triangles,
 denoted by $T_j(\mathbf{v}_i)$, $j=1,\ldots,N_n(\mathbf{v}_i)$.
The shaded area around each vertex $\mathbf{v}_i$  defines the ``barycenter control area'',
which is the area associated to each vertex 
(or 3D $\mathbf{k}$-space vector on the Fermi surface). The control area ($S_i$)
 of a vertex $\mathbf{v}_i$ is calculated by considering the barycenter 
of the neighbouring triangles and the middle points of the vectors connecting the neighbouring triangles, hence the name. This area is the sum of $\frac{1}{3}$ of each neighbouring triangle area,
\begin{eqnarray}\label{eq:bariarea}
S_i = \sum_{j=1,N_n(\mathbf{v}_i)} A[T_j(\mathbf{v}_i)]/3,
\end{eqnarray}
where $A[T_j(\mathbf{v}_i)]$ denotes the area of the nearest neighbour triangles  $T_j(\mathbf{v}_i)$ 
of vertex $\mathbf{v}_i$. This barycenter control area is the simplest 
possible choice and provides a very simple quadrature formula for the integral of a function defined on  the Fermi surface, 
\begin{eqnarray}\label{eq:quadf}
\int_{FS} d^2s_k f(k) \simeq \sum_i S_i f(\mathbf{v}_i).
\end{eqnarray}
It is easily demonstrated that the quadrature formula given by (\ref{eq:quadf}) for the integral
 -and scalar product among functions- is 
absolutely equivalent to the procedure one would obtain by linearly interpolating
 the function $f(k)$ within each triangle
and integrating the linearly interpolated function.
 Thus, (\ref{eq:quadf}) is the generalization of the trapezoidal integration rule
in a -boundary free- surface.  

Once the triangulated surface is constructed, and with all the above information at hand, the discrete 
version of the Laplace operator is numerically available only by identifying the nearest neighbour vertices ($\mathbf{v}_j$) and triangles  of a given vertex $\mathbf{v}_i$. 
In the linear approximation described  above, the Laplace operator comes as a function -only- of the control area $S_i$ and the two -opposite- angles, 
$\alpha_{i,j}$ and $\beta_{i,j}$, of the triangles sharing the edge joining the vertices $\mathbf{v}_i$ and $\mathbf{v}_j$ (see figure~\ref{fig:baric}). Thus, 
the two point centered formula for the second derivative in one dimension is generalized 
by \cite{LB-Mayer,LB-Ch}  
\begin{eqnarray} \label{eq:disc-lap}
\nabla_{k}^2 f(\mathbf{k}) |_{\mathbf{k}=\mathbf{v}_i}  \simeq - \frac{1}{S_i} \sum_{j=1,N_n(\mathbf{v}_i)} \Omega_{i,j} f(\mathbf{v}_j),
\end{eqnarray}
where
\begin{eqnarray}  \label{eq:angles-ab}
\Omega_{i,j} &=&\left\{ \begin{array}{rl}
  -\frac{1}{2} \left [ \mbox{cot} \left (\alpha_{i,j} \right ) + \mbox{cot} \left (\beta_{i,j} \right ) \right ]
 &\mbox{ $i\ne j$} \\
  \sum_{i \ne j} \Omega_{i,j} \ \ \ \ \ \ \ \ \ \  \ \ \ &\mbox{ $i= j$ }.
       \end{array} \right.
\end{eqnarray}

Thus, since only nearest neighbour vertices contribute to
 (\ref{eq:disc-lap}) and (\ref{eq:angles-ab}),
the discretized version of the Helmholtz equation for the HFSH  and the BHFSH set are given, respectively, by the following two 
 generalized -highly sparse- eigenvalue problems:
\begin{eqnarray}\label{eq:HE}
\frac{v (\mathbf{v}_i)}{S_i } \sum_j \Omega_{i,j} \; \Phi_L^j  &=& \omega_L \Phi_L^i,  
\end{eqnarray}
and
\begin{eqnarray}\label{eq:HEb}
\ \ \ \ \frac{1}{S_i } \sum_j \Omega_{i,j} \; \Psi_L^j  &=&  {\kappa}^2_L \Psi_L^i. 
\end{eqnarray}

Note that if the area of the control cells of all vertices 
($S_i$) were equal and the velocity [$v(\mathbf{v_i})$] was constant in (\ref{eq:HE}), 
we would then have a regular eigenvalue problem for the
eigenfunctions $\Phi_L$ and the eigenvalues $\omega_L$ which are labelled by $L$.
The same applies to (\ref{eq:HEb}).
Moreover, since the $\Omega_{i,j}$ matrix is symmetric and the $\frac{S_i}{v (\mathbf{v}_i)} \delta_{i,j}$ operator is  positive definite,
the reality of the eigenvalues is automatically guaranteed. 
%
%
The linear problems in  (\ref{eq:HE}) and (\ref{eq:HEb}) are solved with the aid of the FEAST sparse eigenvalue solver library \cite{FEAST}.

\subsection{Periodic boundary conditions}
As mentioned, in our computational scheme 
a first step consists on finding the list of triangles sharing a given vertex of the triangulation. 
Periodic boundary conditions are implemented by imposing that all the vertices located at boundary planes of the BZ 
and differing by a reciprocal lattice vector $\mathbf{G}$, share the same triangular simplexes. 
Thus, the setup of the periodic boundary conditions implies not only 
accounting for all triangles shared by a point $\mathbf{k}$ (in a given boundary plane) but also adding to this list
those triangle simplexes  which are neighbouring an equivalent boundary plane and sharing the 
vertex $\mathbf{k+G}$.
When a given vertex is located at two boundary planes at the same time (an edge of the BZ) the same
 procedure described above is imposed twice.

\subsection{Fermi surface relaxation: Refinement of the triangular mesh by application of the Newton-Raphson method.} \label{sec:newton}

Within the marching tetrahedra algorithm (\ref{marchingt}), the entire Brillouin zone is sampled by a tetrahedral subdivision and
the energy bands are explicitly calculated only at the corners of the tetrahedral simplexes. 
If a given tetrahedral simplex is found entirely above (or below) the Fermi level, the tetrahedron in question
does not  intersect the Fermi surface. 
When the corners of a tetrahedral simplex lie at both sides
of the Fermi level,
the band energies are linearly interpolated along the edges of the tetrahedron,
 allowing for a linear
estimation of the positions of the three (or four) corners defining the intersection
 of the Fermi surface and the tetrahedral simplex.
In this way, the initial 3D manifold is reduced to a 2D one.

Clearly, if the initial 3D mesh is relatively coarse or/and if the contribution of the 
second or higher order derivatives of the electron energy are appreciable, the error in the determination of the Fermi vectors are substantial.
Once the 2D surface triangulation is obtained by the marching tetrahedra algorithm, we include a second step 
where the 2D $\mathbf{k}$ vectors are allowed to move along the normal to the Fermi surface such that the relaxed vector -and up to numerical precision- lies exactly at the Fermi surface.
In this way, we iteratively improve the quality of the mesh by application of a generalized version of the Newton-Raphson type 
algorithm represented by the following iterative formula,
\begin{eqnarray}\label{eq:Newton}
\mathbf{k}_{(n+1)} =  \mathbf{k}_{(n)} - {\boldsymbol{\nabla}} \epsilon_{\mathbf{k}_{(n)}} \frac{ \left ( \epsilon_{\mathbf{k}_{(n)}}-e_F \right)}
{\left | {\boldsymbol{\nabla}} \epsilon_{\mathbf{k}_{(n)}}\right |^2}.
\end{eqnarray}
In (\ref{eq:Newton}) 
the electron band indices are not shown
 for simplicity, but this relaxation scheme is applied to
all $\mathbf{k}$ triangle vertices defining the Fermi surface and all bands composing the Fermi surface.

Note that the application of (\ref{eq:Newton}) is only affordable due to the low computational cost of the electronic
band energies and velocities (energy gradients) through the Wannier method (see \ref{sec:wannier}).
All the cases that have been investigated have shown that 
5 to 7 iterations ($n$)  are more than sufficient to determine the Fermi wave vectors up to -our- numerical precision $\sim$$10^{-7}$~eV. 

The application of the Newton-Raphson method as introduced above, improves the mesh quality and the accuracy
of the eigenvalue problems represented by equations (\ref{eq:HE}) and (\ref{eq:HEb}),
but this additional step is not 
 essential in our numerical scheme and may be ignored if a high quality triangulation
is obtained by any other method. 

\subsection{Density of states (DOS)}
Obviously, the refinement of the mesh improves the estimation of Fermi surface integrals in any theory or approach. Moreover, several important 
physical quantities involve the integral of the inverse of the velocity (energy gradient), the simplest of which being the density of states (DOS)
$\rho(E)$,
\begin{eqnarray}\label{eq:DOS}
\rho(e_F) = \frac{2}{\Omega_{BZ}} \sum_{n} \int \frac{d^2s_{\mathbf{k},n}} {v_{\mathbf{k},n}} 
\end{eqnarray}
where $d^2s_{\mathbf{k},n}$ is the infinitesimal surface element  and {$n$} denotes the electron band index (the $2$ pre-factor assumes spin degeneracy). 
We approximate the integral in (\ref{eq:DOS}) by considering the barycenter control areas ($S_{i,n}$) obtained by triangulation
and explicitly calculating the velocities [$v_n(\mathbf{v}_i)$] for all the relaxed vertices $i$ and bands $n$,
\begin{eqnarray}\label{eq:DOSn}
\rho(e_F)
\simeq \frac{2}{\Omega_{BZ}} \sum_n \sum_i \frac{S_{i,n}}{v_n(\mathbf{v}_i)}.
\end{eqnarray}

In the linear tetrahedron method \cite{tetrahedrom} 
 (\ref{eq:DOS}) is treated -in essence- by considering the
electron velocities and vertex positions linearly interpolated in the inner volume of each tetrahedron.

In our numerical scheme two improvements are introduced for computing the integral in
 (\ref{eq:DOS}):
(i) the electron velocities are explicitly calculated for all the 2D surface $\mathbf{k}$ point vertices and 
(ii) the triangular mesh is iteratively improved by forcing the triangle vertices to 
lie at the Fermi surface.
Thus, the above method for estimating Fermi integrals  may be considered as a surface specialized
non-linear version of the linear tetrahedron method.

\subsection{Topological characterization: Euler characteristic and the Gauss-Bonnet theorem applied to Fermi surfaces}

The Gauss-Bonnet theorem is one the most prominent theorems in differential geometry connecting a local property of the surface
such as the Gaussian curvature, and a global topological property such as the genus or the Euler characteristic. 

The Fermi surface is a compact and periodic surface and does not present any boundary. With these restrictions,
the Gauss-Bonnet theorem is stated as follows: For a given Fermi sheet $n$ and $K(\mathbf{k})$ being the Gaussian curvature at point $\mathbf{k}$ of the surface, 
the Euler characteristic $\chi$ is given by
\begin{eqnarray}\label{eq:GaussB}
\frac{1}{2 \pi}\int d^2s^n_{\mathbf{k}} \,\, K(\mathbf{k}) = \chi^n.
\end{eqnarray}

We have implemented the above formula as a quality test of our approach and we have checked that the result is completely 
independent of the choice of the Brillouin zone.

Numerically, the Gaussian curvature is given by 
\begin{eqnarray}\label{eq:angledefect}
K(\mathbf{k})|_{\mathbf{k}=\mathbf{v}_i} \simeq   \frac{\left( 2\pi - \sum_{j=1,N_n(\mathbf{v}_i)} \tau_{i,j} \right )}{S_i},
\end{eqnarray}
where $\tau_{i,j}$ denotes the angle between the triangle edges joining 
vertices $\mathbf{v}_i$ and $\mathbf{v}_j$, and $\mathbf{v}_i$ and $\mathbf{v}_{j+1}$
(see figure~\ref{fig:baric}) and
$S_i$ is defined in (\ref{eq:bariarea}). Thus, 
one obtains that numerically (\ref{eq:GaussB}) can be rewritten as
\begin{eqnarray}\label{eq:Descartes}
\sum_{i}  \left( 1 - \frac{1}{2\pi} \sum_{j=1,N_n(\mathbf{v}_i)} \tau_{i,j} \right ) = \chi,
\end{eqnarray}
which is the exact Descartes theorem on total angular defect of a polyhedron. 

The genus of a single-sheet surface is given by  
\begin{eqnarray}\label{eq:genus}
g = 1-\chi/2.
\end{eqnarray}
While $\chi$ is additive for multiple-sheet surfaces, $g$ must be considered separately for each sheet.
It is worth to mention that we obtain the above topological numbers by a direct application of (\ref{eq:Descartes}), and that
we obtain an integer number up to our numerical accuracy. 

Even though the genus of several Fermi surfaces such as Lithium 
are trivial  ($g=0$), the same is not true even for a  free-electron-like material such as Cu, where we find $\chi=-6$ (and $g=4$).

\subsection{General properties of the HFSH set}\label{propertiesHFSH}

Although Allen's FSH \cite{FSH-Allen} and the HFSH presented in this article are
{defined} very differently  in the sense that
the FSH are constructed as  explicit orthogonal polynomials and 
the HFSH set is generated  as a solution of the Helmholtz equation defined on the Fermi surface 
[(\ref{eq:helmholtz}) and (\ref{eq:HE})], both sets present very important similarities
in their properties.
First, being a solution  of a second order differential equation, the HFSH set is orthogonal; and second, our choice for the normalization in (\ref{eq:ortHFSH}) is deliberately  chosen equal to that introduced in \cite{FSH-Allen}. 

Allen rewrote the anisotropic Boltzmann and Eliashberg equations in terms of the polynomial FSH set.
One concludes that as the scalar product of the HFSH set is defined exactly as for the FSH, the expressions derived by Allen for the Boltzmann transport and
the anisotropic superconductivity in terms of FSH are still valid  for the HFSH set presented in this work. 
This is the reason why in this article we concentrate on the calculation and description of the HFSH set and 
we refer to  \cite{FSH-Allen} and \cite{FSH-Allen-tr} for a detailed treatment of the Boltzmann and Eliashberg equations
using FSH. In the next lines we face the problem of function representation in terms of HFSH.  

The HFSH set allows us to efficiently represent any function defined on  the Fermi surface, but more important,
the HFSH modes allow to express integro-differential equations involving quasi-elastic scattering processes much more economically
and probably in a physically more transparent way.
Good examples of anisotropic functions defined on the Fermi surface are for instance the electron lifetime, $\tau(\mathbf{k})$, the electron 
mass enhancement, $\lambda(\mathbf{k})$, the superconducting gap, $\Lambda(\mathbf{k})$, electron velocity components, etc.
Similar to any integral transformation, the smoother the function to be represented the more efficient is the HFSH representation method.
In the next subsections, we formalize the problem of functional representation considering the HFSH modes.  

Let us consider any of these physical properties as a generic function, $F(\mathbf{k})$, 
and let us  rewrite this function in terms of the HFSH set,
\begin{eqnarray}\label{eq:CL}
F (\mathbf{k}) = \sum_{L} c_L(F) \Phi_L (\mathbf{k}).
\end{eqnarray}
We refer to (\ref{eq:CL}) as the HFSH representation of the $\mathbf{k}$ dependent function 
$F(\mathbf{k})$ defined on the Fermi surface.
The scalar product and normalization of the HFSH $\left\{\Phi_L \right\}$ are defined by (\ref{eq:ortHFSH})
in momentum space,
thus the components $c_L(F)$ of the expansion become directly accessible, 
the coefficients of the expansion having the same units as the original $F(\mathbf{k})$ function,
\begin{eqnarray}\label{eq:CLint}
c_L(F) = \langle \Phi_L | F  \rangle \equiv
\frac{\int \frac{d^2s_{\mathbf{k}}}{v(\mathbf{k})} \Phi_L (\mathbf{k}) F (\mathbf{k})}{\int \frac{d^2s_{\mathbf{k}}}{v(\mathbf{k})}}.
\end{eqnarray}

Of course, as the HFSH set is orthogonal, the scalar product of two functions $F_1$ and $F_2$ may be written
conveniently in terms of the HFSH components of these functions, 
\begin{eqnarray}
\langle F_1 | F_2 \rangle = \sum_L c_L(F_1) c_L(F_2) = 
\frac{\int \frac{d^2s_{\mathbf{k}}}{v(\mathbf{k})} 
F_1(\mathbf{k}) F_2(\mathbf{k})}{\int \frac{d^2s_{\mathbf{k}}}{v(\mathbf{k})}}. \ \ \ 
\end{eqnarray}

The procedure of transforming a function $F(\mathbf{k})$ into a discrete set of 
coefficients $c_L(F)$ is very similar to ordinary
Fourier transformation or the spherical harmonics expansion in the unit sphere.
For sufficiently well behaved functions we should expect that a relatively small 
amount of HFSH modes is enough but, in any case,
one has the control on the desired accuracy by tuning the energy cutoff. 

The product of two functions $F_1(\mathbf{k})F_2(\mathbf{k})$ defined in $\mathbf{k}$ space may be represented
as a function of the HFSH components  of each of these functions 
separately [$c_L(F_1)$ and $c_L(F_2)$] with the following relation that generalizes the convolution theorem 
in Fourier analysis 
\begin{eqnarray}\label{eq:multi}
F_1(\mathbf{k}) F_2(\mathbf{k}) =\!\! \sum^L_{L_1,L_2} \Xi_{L;L_1,L_2}\,\, c_{L_1} (F_1) \,\, c_{L_2} (F_2)\,\, \Phi_L(\mathbf{k}), \ \ \ \
\end{eqnarray}
where the matrix elements $\Xi_{L;L_1,L_2}$ play the same role as the 
Clebsch-Gordan coefficients for ordinary spherical harmonics,
\begin{eqnarray}\label{eq:xi}
\Xi_{L;L_1,L_2} = \langle \Phi_L |  \Phi_{L_1} \Phi_{L_2}  \rangle. 
\end{eqnarray}
The coefficients $\Xi_{L;L_1,L_2}$ are symmetric with respect to the 
permutations of the $L,L_1,L_2$ indices, as {it} was also found by Allen for the FSH set \cite{FSH-Allen}.
Some elementary properties of $\Xi_{L;L_1,L_2}$ are (the first element of the HFSH set is $L=1$)
\begin{eqnarray}
\nonumber
\Xi_{L;1,1} &=& \delta_{L,1}  \ \ \  \Xi_{1;L_1,L_2} = \delta_{L_1,L_2}. 
\end{eqnarray}

The $\Xi$ coefficients allow to expand the product of two $\Phi$ functions in terms of a simple linear combination of 
HFSH elements and vice-versa,
\begin{eqnarray}
\nonumber
\Phi_L(\mathbf{k}) =\sum_{L_1,L_2} \Xi_{L;L_1,L_2} \Phi_{L_1}(\mathbf{k}) \Phi_{L_2}(\mathbf{k})
\\
\Phi_{L_1}(\mathbf{k}) \Phi_{L_2}(\mathbf{k}) = \sum_{L} \Xi_{L;L_1,L_2}  \Phi_L(\mathbf{k}).
\end{eqnarray}

The scalar product of pairs of HFSH {may} 
be expressed also in terms of the $\Xi$ matrix elements, 
\begin{eqnarray}
\langle \Phi_{L_1} \Phi_{L_2}|\Phi_{L_2} \Phi_{L_3} \rangle =  \sum_{L} \Xi_{L;L_1,L_2} \Xi_{L;L_2,L_3}.
\end{eqnarray}

In principle, one would calculate all the $\Xi_{L;L_1,L_2}$ coefficients only once and try 
to reduce all redundancies as much as possible. 

Let us now consider the HFSH representation  of a matrix element of any physical magnitude 
with two momentum indexes.
The electron-phonon matrix elements $g_{\mathbf{k},\mathbf{k}'}$ as well as 
many other physical quantities, such
as the non-local self-energy or the impurity scattering matrix elements, need to be represented generally in terms of a pair on electron momenta $\mathbf{k}$  and $\mathbf{k}'$. 
In all these cases we would follow a similar procedure,
\begin{eqnarray}\label{eq:gCL}
g_{\mathbf{k},\mathbf{k}'} = \sum_{L,L'} c_{L,L'}(g) \Phi_L (\mathbf{k}) \Phi_{L'} (\mathbf{k}'),
\end{eqnarray}
where
\begin{eqnarray}\label{eq:dgCL}
 c_{L,L'}(g) = 
\frac
{\int \frac{d^2s_{\mathbf{k}}}{v(\mathbf{k}) } \frac{d^2s_{\mathbf{k}'}}{v(\mathbf{k}') } \ g_{\mathbf{k},\mathbf{k}'} \  \Phi_L (\mathbf{k})\Phi_{L'} (\mathbf{k}') }
{\left (\int \frac{d^2s_{\mathbf{k}}}{v(\mathbf{k})}\right )^2}.
\end{eqnarray}
Of course the function $g_{\mathbf{k},\mathbf{k}'}$ should be reasonably smooth for a good quality representation in terms of the HFSH as described above.
When the quantity in question is a scattering amplitude or a complex matrix element,
 one must first fix the complex arbitrary phases 
of $g_{\mathbf{k},\mathbf{k}'}$.
 If  time reversal and inversion symmetries are both present, these  phases are easily 
removed \cite{wannier-original}, but more generally, one is forced to fix these phases by a Wannier procedure
 \cite{wannier-original,EigurenPRB08,GiustinoPRL}.
Alternatively, one could consider the absolute values of the  matrix elements
 $|g_{\mathbf{k},\mathbf{k}'}|^2$.

The above algebraic machinery has a great potential in restating integro-differential problems
defined on the Fermi surface. We refer to  \cite{FSH-Allen,AllenMitrovic} for a detailed 
description of the procedure for transforming the Boltzmann transport and the Eliashberg theory for the FSH, which would 
be completely valid for our HFSH set.

\subsection{Numerical tabulation of an anisotropic quantity}

One of the main problems encountered when trying to characterize an anisotropic physical quantity defined 
on  the Fermi surface, is that 
the only accessible method is a graphical representation through a colour code,
which requires the computation of that quantity on a large amount (of the order of $10^5$) of  $\mathbf{k}$ vectors
defining the FS. However, this method is mainly visual and
not quantitative. 

The representation of a function $F(\mathbf{k})$ using the HFSH expansion coefficients $c_L(F)$ enables a direct 
quantitatively description of the anisotropy. 
Indeed, the HFSH method allows to tabulate numerically any complex anysotropic quantity by means of about 10$^2$ coefficients,
the most representative being those corresponding to the modes with the lowest energy.

Since $\Phi_{L=1}(F)=1$, the first expansion coefficient, $C_{L=1}(F)$, yields  directly 
the FS average of the function $F(\mathbf{k})$. In materials with spherical or nearly spherical symmetry,
such as Li or Cu (see section \ref{freeLiCu}), the next three modes [$\Phi_{L}(F), L=2,3,4$] are very similar to the 
$p_x$,  $p_y$ and $p_z$ spherical harmonics.
In general, we find that a few number ($\sim 20$) of coefficients is sufficient for capturing the most significant part of the
anisotropy of any $\mathbf{k}$ dependent function.

In this sense, the numerical tabulation appears to be a potentially important application of the HFSH set. Let us 
 consider as a standard example, the anisotropic electron-phonon mass enhancement, $\lambda_{\mathbf{k}}$, 
or the momentum dependent lifetime, $\tau_{\mathbf{k}}$. 
Following a standard procedure, if one wanted to compare two different calculations of  $\lambda_{\mathbf{k}}$, for instance,
obtained using two different computation methods, one would need to compare the  $\mathbf{k}$ dependent 
data set point by point. This is, obviously, not practical and results on a high symmetry direction might be practically checked, if at all.

In a HFSH mode expansion a list of a few numerical coefficients [$c_{L}(F)$] would be
 sufficient  to compare, at least, the grossest details
of the directional dependence of any magnitude, and the tabulation of any anisotropic quantity would then be easily accessible.

\subsubsection{Denoising,  filtering and mismatch error analysis}\label{filtering}

The HFSH is a complete set and the finest details  
of  any quantity are accessible only by increasing the cutoff energy (up to the triangular grid capability).
However, it may happen that a quantity calculated on  the Fermi surface is accompanied by a noisy background, which is
a situation that could be standard experimentally.

Let us suppose that we have calculated 
a physical property  $F({\mathbf{k}})$ explicitly for all the vertex $\mathbf{k}$ points of our triangular grid describing the FS.
Consider a finite cutoff ($E_c$) for  the expansion of $F({\mathbf{k}})$ in terms of the HFSH set
using  (\ref{eq:CL}):
\begin{eqnarray}\label{eq:CLrepr}
\tilde{F}^{N_L}(\mathbf{k}) = \sum^{N_L}_{L=1} c_L(F) \Phi_{L}(\mathbf{k}),
\end{eqnarray}
where $N_L$ denotes the number of modes such that $\omega_L~<~E_c$.
We then obtain a smoothed  function $\tilde{F}^{N_L}(\mathbf{k})$ where the
 fine details smaller than a wave length
\begin{eqnarray}
\lambda_c \sim 2 \pi \sqrt{\frac{C_{L=1}}{\omega_L}} 
\end{eqnarray}
are filtered.

A measure of the mismatch error when considering only a finite set of HFSH modes, $N_L$ might be
 obtained by the Fermi surface integral 
\begin{eqnarray}\label{eq:missm}
\epsilon(N_L) = \frac{\int  \frac{d^2s_\mathbf{k}}{v_{\mathbf{k}}} \left | F(\mathbf{k}) - \tilde{F}^{N_L}(\mathbf{k})  \right | }
{ \int  \frac{d^2s_\mathbf{k}}{v_{\mathbf{k}}} \left | F(\mathbf{k}) \right | }, 
\end{eqnarray}
which is approximated by a simple linear quadrature formula as in (\ref{eq:quadf}).

\section{Test examples for real materials}
 
We have considered materials with one Fermi sheet (bcc-Li, bcc-Na and fcc-Cu),
 with two Fermi sheets 
(fcc-Pb and bcc-W) and with three Fermi sheets (hex-MgB$_2$) as testing examples. 

The DFT ground state for all of them have been obtained using norm-conserving pseudopotentials
 with the PBE \cite{PBE} functional and with a cutoff energy of 30~Ry for bcc-Li, 
50~Ry for bcc-Na and fcc-Pb, 55~Ry for bcc-W, 60~Ry for hex-MgB$_2$, and 110~Ry for fcc-Cu.
Each DFT ground state was in turn used as an input for the Wannier calculations which were 
carefully converged.

\begin{table}[h]
\caption{\label{tab:table1}
Density of states, area of the Fermi surface, Euler characteristic and genus obtained using the method
presented in this work for different examples with one Fermi sheet (bcc-Li, bcc-Na and fcc-Cu), two Fermi sheets (fcc-Pb and bcc-W) and three Fermi sheets (hex-MgB$_2$). The area, the Euler characteristic and the genus are shown for each of the Fermi sheets.
The second column shows the number of kpoints used in the Wannier calculation
 (see  \ref{sec:wannier}). The last 
column shows the density of states obtained from the ground state calculation done with Quantum Espresso \cite{espresso} using the linear tetrahedron method \cite{tetrahedrom} using a grid in 
momentum space of $51^3$ in all cases.
}
\begin{indented}
\lineup

\item[]\begin{tabular}{l c c c c c c c}
\br
&\textrm{$\mathbf{k}$ grid} & \textrm{DOS [\eref{eq:DOS}]} & \textrm{Area} & $\chi$ & $g$ & \textrm{DOS (Linear tetrahedron)} \\
& & \textrm (states/eV) & \textrm{(($2\pi/a$)$^2$)} &  & & \textrm{(states/eV) } \\
\mr

\multirow{2}{*}{Li} &20$^3$ & 0.495 & 4.911 & 2 & 0 &\multirow{2}{*}{0.490}\\
 		    &40$^3$ & 0.492 & 4.886 &2&0\\
\mr

\multirow{2}{*}{Na} &20$^3$ & 0.466 & 4.848 & 2 &  0 &\multirow{2}{*}{0.460}\\
 		    &40$^3$ & 0.465 & 4.849 &2 &0 \\
\mr

\multirow{2}{*}{Cu} &20$^3$ & 0.292 & 7.617 & -6 &  4 &\multirow{2}{*}{0.291}\\
 		    &40$^3$ & 0.291 & 7.597 & -6 &  4 \\
\mr
\multirow{4}{*}{Pb} &\multirow{2}{*}{20$^3$} &\multirow{2}{*}{0.513} & 4.575 & 2 & 0 &\multirow{4}{*}{0.506}\\
                    &       &       & 7.667 & -12 & 7 &\\
\cline{2-6}
 		    &\multirow{2}{*}{40$^3$} &\multirow{2}{*}{0.510} & 4.577 &2 & 0 \\
 		    &        &       & 7.596 &-12 & 7 \\
\mr
\multirow{4}{*}{W} &\multirow{2}{*}{20$^3$} &\multirow{2}{*}{0.371} & 2.073 & 2$\times$(6+1) & 7$\times$0 &\multirow{4}{*}{0.383}\\
                    &       &       & 2.330 & 2 & 0 &\\
\cline{2-6}
 		   &\multirow{2}{*}{40$^3$} & \multirow{2}{*}{0.373} & 2.137 & 2$\times$(6+1)&7$\times$0 \\
                    &       &       & 2.316 & 2 & 0 &\\
\mr
\multirow{6}{*}{MgB$_2$} &\multirow{3}{*}{20$^3$} &\multirow{3}{*}{0.698} & 0.531 & 0 & 1 &\multirow{6}{*}{0.703}\\
 		    & & & 1.834 &0 \mbox{and} -2 & 1 \mbox{and} 2\\
 		    & & & 1.664 &-2 & 2\\
\cline{2-6}
 		    &\multirow{3}{*}{40$^3$} &\multirow{3}{*}{0.700} & 0.533 &0 & 1 \\
 		    & & & 1.840 &-2 & 2\\
 		    & & & 1.665 &-2 & 2\\
\br
\end{tabular}
\end{indented}
\end{table}

Table~\ref{tab:table1}  summarizes the density of states, area of the Fermi surface,
 Euler characteristic and genus for all the examples. The second column shows the 
 momentum sampling for the tetrahedral division in the marching tetrahedron
method for constructing the triangulation of the Fermi surface.
 The third column displays the  DOS obtained using the method presented in this work
[using (\ref{eq:DOS})], which includes a relaxation of the
 surface, as introduced in section \ref{sec:newton}, and an explicitly calculated velocity 
for each vertex position of the triangulation
using the Wannier method, as explained in \ref{sec:wannier}. These
results are compared to the  linear tetrahedron method \cite{tetrahedrom} (last column), 
where both, the vertex
(without relaxation) and the magnitude of the velocity are included effectively by linear direct interpolation.
The agreement between both values is remarkable in all cases and allows us to  check that
 the quadrature formula 
for the integrals is accurate when a triangular division together with a barycenter 
control cell (see figure~\ref{fig:baric}) is considered.

The fourth column shows the calculated area of the different Fermi sheets for each momentum space sampling.
The next two columns show, for each Fermi sheet,
 the calculated Euler characteristic ($\chi$) and genus ($g=1-\chi/2$ for a connected surface).
As a convergence test, all these magnitudes were evaluated for the two different sampling densities, 20$^3$ and 40$^3$. 
Note that all the quantities are found to be practically converged already for a sampling density  of 20$^3$.

Bcc-Li and bcc-Na present a Fermi surface which is topologically trivial and this is confirmed by a value of the  Euler characteristic of $\chi$=2 ($g$=0) in both cases.  
The topological classification of the Fermi surfaces is not the goal of this work,
 but since internal angles of the triangles simplexes enter 
(\ref{eq:disc-lap}) and (\ref{eq:angles-ab}) in a way which is crucial, 
a direct computation of the Gaussian 
curvature, considering the total angular defect in (\ref{eq:angledefect}),
 is an essential test of consistency.
Regardless of the choice of the unit cell, the tetrahedral sampling, or the complexity of the surface,
we obtain in all cases that the Euler characteristic ($\chi$) is recovered as an integer number up to double real numerical precision ($\sim10^{-13}$). 
As an application of a more complex example, we mention the Euler characteristic of the first band of bcc-W [$\chi=2\times (6+1)$],
 which can be easily understood by inspecting its Fermi surface (not shown).
It is composed of 6 disjoined ellipsoidal sheets around the high symmetry point $N$, and  one diamond-shaped sheet centered at $H$,
 each of them yielding a genus $g=0$.

As for the visualization of the HFSH,  we have considered only bcc-Li,  fcc-Cu and 
hex-MgB$_2$ as representative examples.
Bcc-Lithium and fcc-Cu are found to be reasonably close to the ordinary spherical harmonics, although, the Fermi surface of fcc-Cu is topologically not so trivial as that of  bcc-Li. 
The hex-MgB$_2$ Fermi surface presents three different electron bands crossing the Fermi level,
with one of these bands composed by two disconnected pieces. 
MgB$_2$ enables us to demonstrate the utility of the method 
in a more complex situation, but on the same footing as in simpler Fermi surfaces.

\subsection{Free electron like bcc-Li and fcc-Cu} \label{freeLiCu}

Our first example applications are bcc-Li and fcc-Cu at ambient pressure, which are textbook examples of free-electron-like systems.
The Fermi surface of these materials is approximately spherical and the HFSH functions obtained from (\ref{eq:HE}) [and (\ref{eq:HEb})] 
should be expected to approximate the ordinary spherical harmonics.

\begin{figure} [h]
\centerline{\includegraphics[width=\columnwidth]{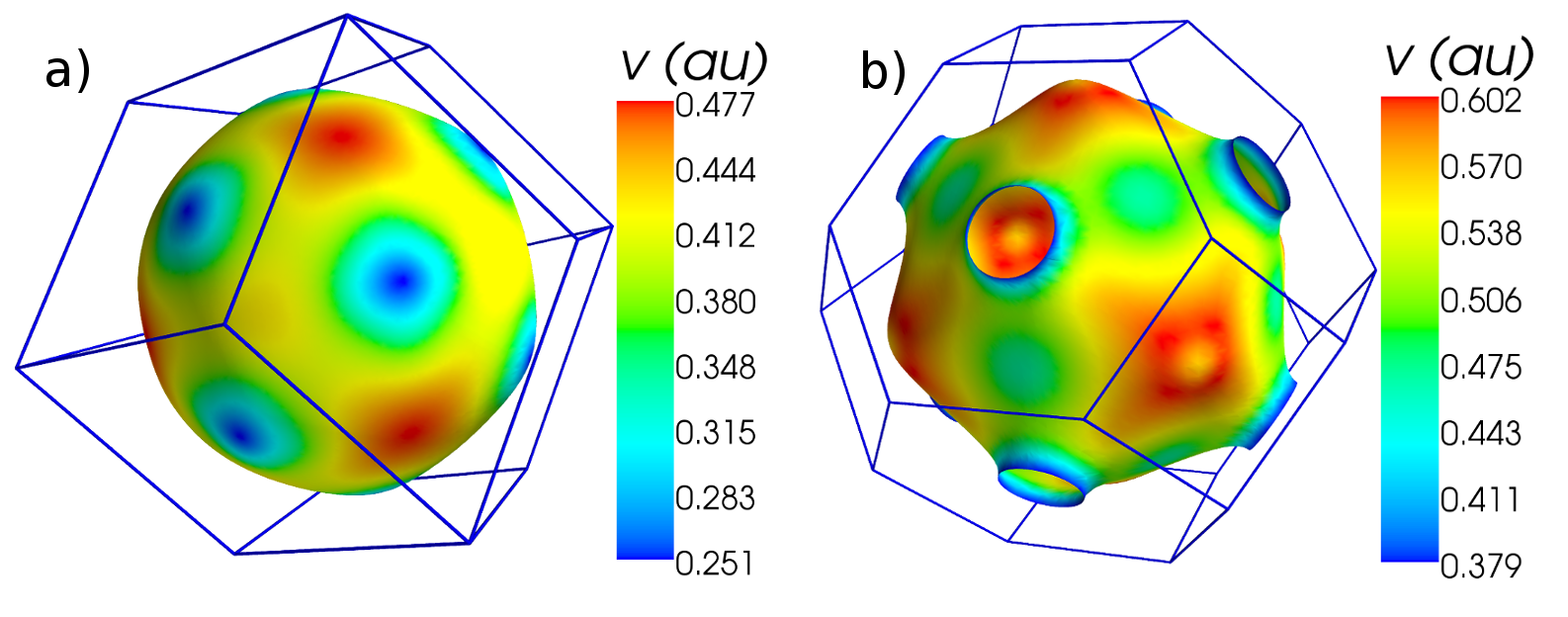}}
\caption{Fermi surface of bcc-Li (a) and fcc-Cu (b) and the modulus of the electron velocity $v_{\mathbf{k}}$ (colour code) as a function of the electron momentum.}
\label{fig:LiCu-tri}
\end{figure}

Figures~\ref{fig:LiCu-tri}a and \ref{fig:LiCu-tri}b show 
the calculated Fermi surface of bcc-Li and fcc-Cu, respectively,
considering a momentum space division of 40$^3$ in both cases. The marching tetrahedron algorithm 
(as introduced in \ref{marchingt}) produced 27962 (bcc-Li) and 24582 (fcc-Cu) triangle vertices
 defining the FS, which
were relaxed until they were located within a window 
of $\left|\epsilon_{\mathbf{k}}-E_F\right |<10^{-6}$~eV. 
%
The colour code in figures~\ref{fig:LiCu-tri}a and  ~\ref{fig:LiCu-tri}b corresponds to the absolute value of the electron 
velocity which enters (\ref{eq:helmholtz}) and (\ref{eq:HE}).
Although both, bcc-Li and fcc-Cu, present an almost spherical Fermi surface, we observe that the anisotropy of the electron velocity is not negligible.

\begin{figure} [h]
\centerline{\includegraphics[width=\columnwidth]{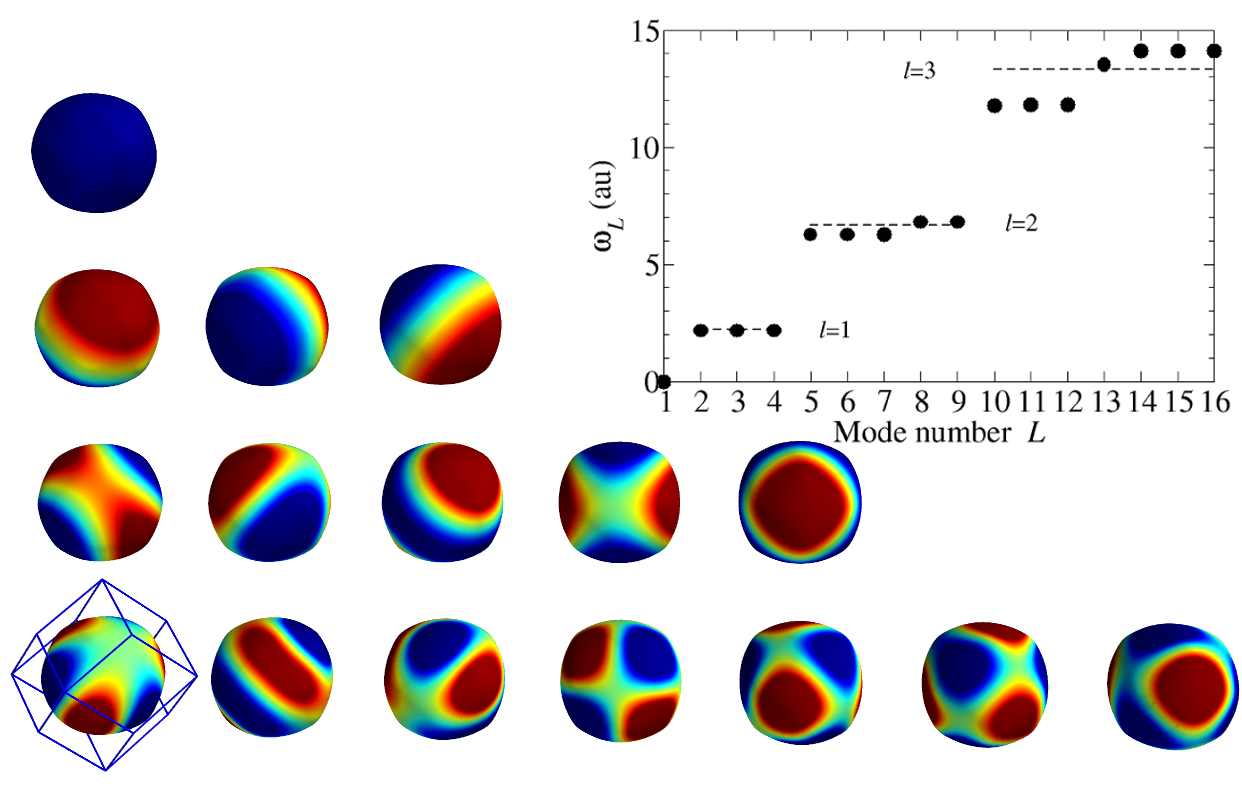}}
\caption{Lowest energy 16 HFSH modes of bcc-Li. The HFSH functions are ordered by rows
 following the degree of  degeneracy [$n_l$=$(2l+1)$] of the 
ordinary spherical harmonics. The inset shows the 
{HFSH} 
 energies ($\omega_L$) for these states 
{(filled circles). The dashed lines represent the 
 energies when (\ref{eq:helmholtz}) is solved analytically in a spherical
surface of radius $k_F$ and average velocity $v_F$, $\omega=l(l+1)v_F/k_F^2$.}
The lowest energy state (first row) is a constant function $\Phi_{L=1}$=$1$, according to
 the -defining- normalization condition in (\ref{eq:ortHFSH}).
 In the second row we find three degenerate HFSH modes,
 approximately resembling the ordinary 
$p_x$, $p_y$, $p_z$ spherical harmonics. In the third and fourth 
rows we show the calculated HFSH modes which are similar to the $d$ and $f$ states,
 but in these case the degeneracy is lifted as perfect spherical
symmetry is absent.}
\label{fig:LiPeriod}
\end{figure}

\begin{figure} [h]
\centerline{\includegraphics[width=\columnwidth]{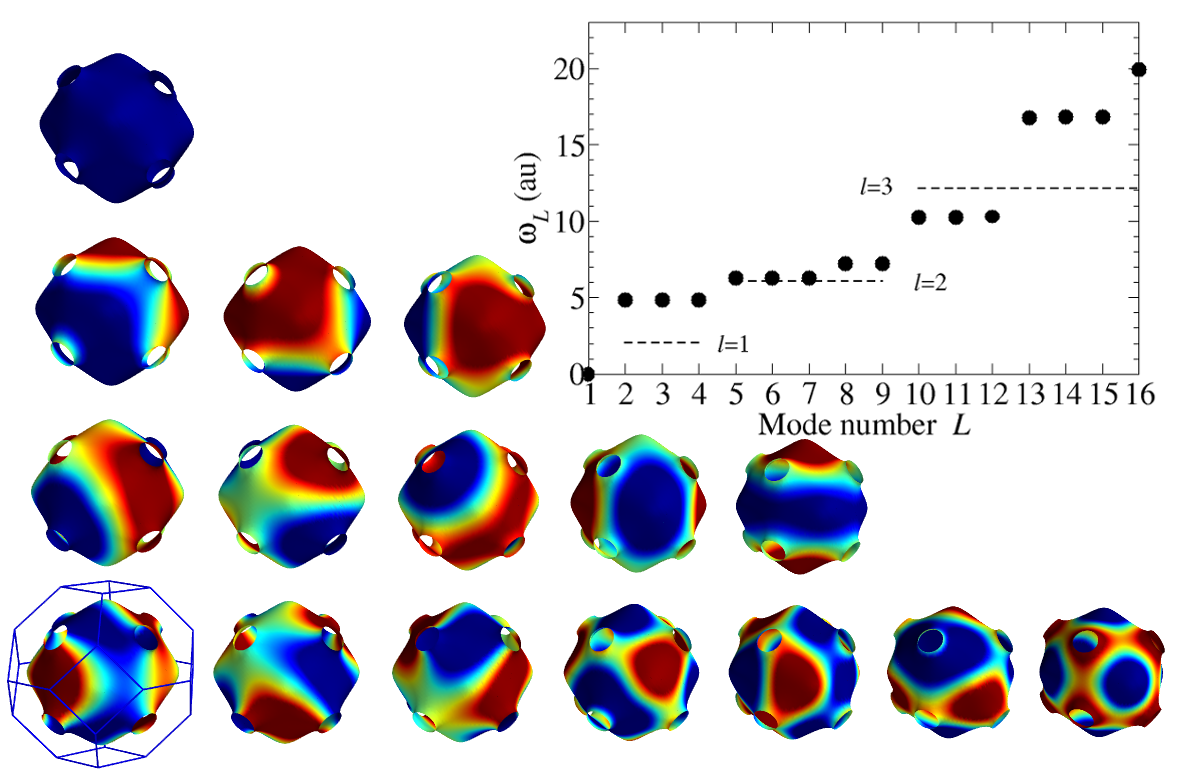}}
\caption{
As in figure~\ref{fig:LiPeriod} for fcc-Cu
}
\label{fig:CuPeriod}
\end{figure}

Figures~\ref{fig:LiPeriod} and \ref{fig:CuPeriod} present the calculated 16 lowest energy HFSH 
modes for bcc-Li and fcc-Cu, respectively. 
The states are shown by following the
same degeneracy ordering as if they were the usual spherical harmonics, i.e., 
the first row corresponds to the constant $s$-like 
state, second one corresponds to the $p_x$, $p_y$, and  $p_z$-like modes,
 third row to the $d$-like set, and so on. The correspondence with the
ordinary spherical harmonics is direct because of the simplicity of this surface. 
The  energy dependence of the HFSH for bcc-Li and fcc-Cu is shown 
 {with filled circles}
in the inset
 of figure~\ref{fig:LiPeriod} and figure~\ref{fig:CuPeriod}, respectively.
We observe that the first non-trivial three states  of $p$-like character (second row)
 are found to be degenerate like the
spherical harmonics. 
However,  the original five-fold degeneracy of the ordinary $d$ spherical harmonics
is broken in both cases generating three degenerate states plus an additional two 
dimensional degenerate subspace. 
Indeed, in these systems the crystal symmetry includes a $p$-like symmetry but not
a $d$-like one and the effect appears as a crystal field effect acting on a spherically
symmetric system.
 Similarly, the seven-fold original degeneracy is lifted in bcc-Li into a set of degenerate 
subspaces of three, one and 
three dimensions respectively (3+3+1 in bcc-Cu). 
For comparison we have also shown in the inset of  figure~\ref{fig:LiPeriod}
and  \ref{fig:CuPeriod} (dashed lines) the degenerate eigenvalues of (\ref{eq:helmholtz})
when the HFSH are taken to be ordinary spherical harmonics 
\begin{eqnarray} \label{eq:ll1KF2}
\omega=\frac{l(l+1) v_F}{k_F^2},
\end{eqnarray}
where $l$ denotes the angular momentum, $v_F$ is the mean velocity
at the FS and $k_F$ is the mean radius of each of the  nearly spherical Fermi surfaces  
displayed in figure \ref{fig:LiCu-tri}. 
Note that in Li (figure  \ref{fig:LiPeriod}), the lowest HFSH energies (filled circles)
reproduce very well the spherical harmonics eigenvalues (dashed lines) for $l=1$ and $l=2$.
Even for $l=3$, the degeneracy of the HFHS eigenvalues is broken
 around the  value given by (\ref{eq:ll1KF2}). 
The lifting of the degeneracy for $l=3$ is stronger in the case of Cu (figure~\ref{fig:CuPeriod}),
where we still find a reasonable agreement between the HFSH energies and
  (\ref{eq:ll1KF2}) for $l=2$.
However, for $l=1$ the ideal spherical harmonics energy deviates from the HFSH eigenvalue,
which we attribute to the fact that the necks of the Fermi surface of Cu represent 
a strong perturbation to the sphere, mostly noticeable at long wavelengths.

Figures~\ref{fig:LiPeriod} and \ref{fig:CuPeriod} demonstrate several general features
of the HFSH functions which are common to those of other more complex systems as well.
The first mode ($L=1$) is a constant function over the surface and its energy is equal to zero.
 This is clear from (\ref{eq:HE}) [and (\ref{eq:HEb}) for BHFHS],
 as the diagonal elements of the Laplace operator are equal to the sum of the rest of the elements in the same row and therefore the summation is zero and any constant function multiplied by the discrete Laplace operator [$\Omega$ in  (\ref{eq:angles-ab})]
 is null. 
Thus, the constant function is always the lowest energy member
 in any HFSH set.

For higher energies the wavelength of the mode oscillations are shorter and the energy increases,
to a very good approximation, linearly with the number of modes. For both, the HFSH and HFSHB sets,
 we find that as the mode number $L \rightarrow \infty $
\begin{eqnarray}\label{eq:linear-HFSH}
\omega_L &\simeq&    \frac{\pi^2 \,\langle v \rangle}{S} L
\end{eqnarray}
and
\begin{eqnarray}\label{eq:linear-BHFSH}
\kappa^2_L &\simeq&   \frac{\pi^2 }{S} L
\end{eqnarray}
Equations (\ref{eq:linear-HFSH}) and (\ref{eq:linear-BHFSH})  work very well already for $L \sim 100$ in all materials -and Fermi sheets- treated in this work. These relations allows us to estimate the number of modes required for a given cutoff energy.

\begin{figure}
\centerline{\includegraphics[width=0.6\columnwidth]{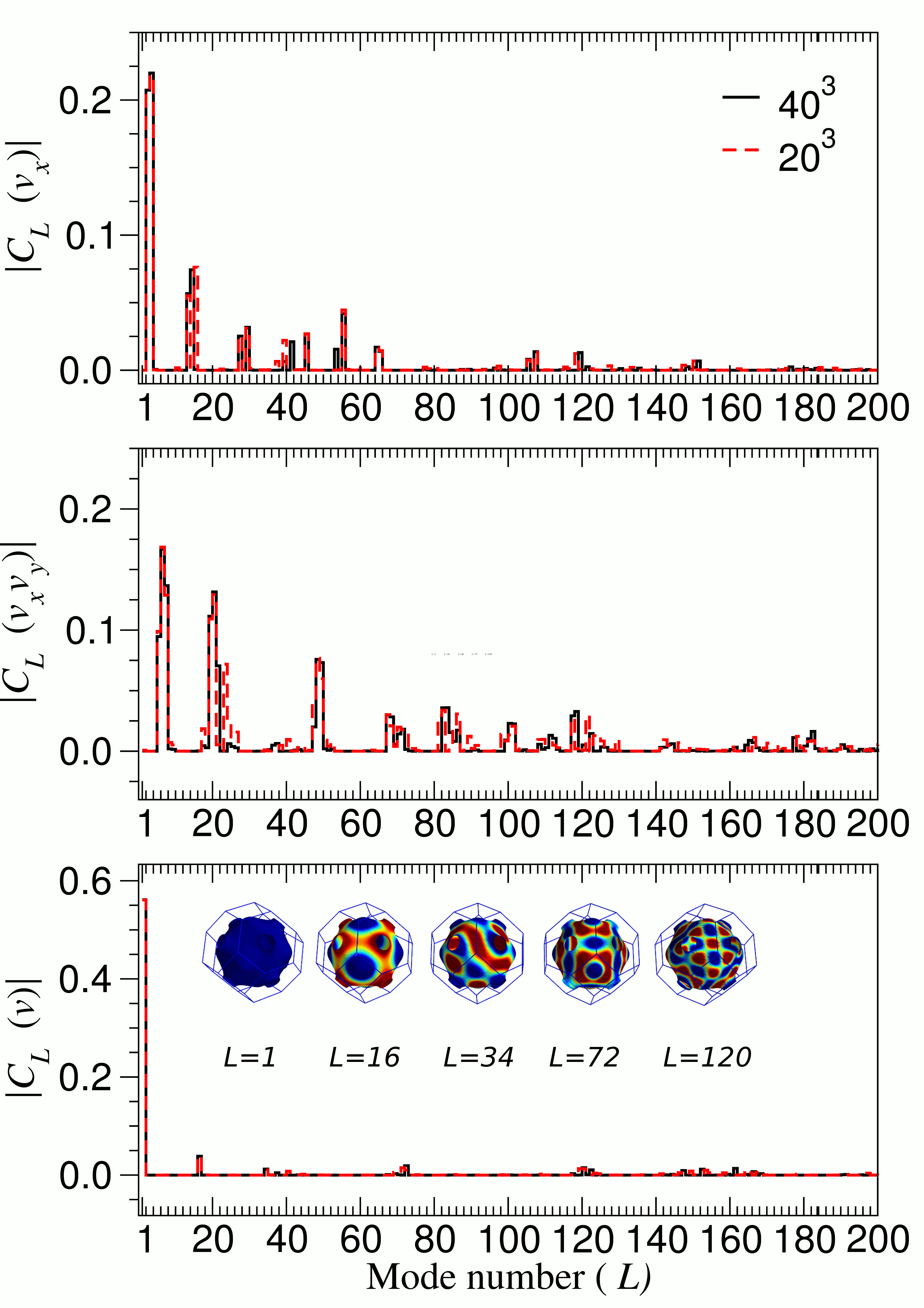}}
\caption{HFSH mode analysis of the cartesian components of the velocity in fcc-Cu.
Absolute value of the $c_L(v)$ components [see (\ref{eq:CL}) and (\ref{eq:CLint})] of the
$x$ component of the velocity [$v_x(\mathbf{k})$] (top),
to the product of  the  $x$ and $y$ components [$v_x(\mathbf{k}) v_y(\mathbf{k}) $] (middle) 
and of the modulus of the electron velocity  [$|v(\mathbf{k})|$] (bottom). 
  The inset shows the  HFSH modes contributing the most to the 
 modulus of the velocity in fcc-Cu.}
\label{fig:CuCL}
\end{figure}

Figure~\ref{fig:CuCL} shows the application of the HFSH mode analysis to the cartesian 
velocity component $v_x(\mathbf{k})$ (top),
the multiplication of the $x$ and $y$ components, $v_x(\mathbf{k})v_y(\mathbf{k})$,
 (middle) and the modulus of the velocity $|v(\mathbf{k})|$ (bottom), 
defined on  the Fermi surface of fcc-Cu. 
We show the absolute value of the coefficients $c_L$,  defined in 
(\ref{eq:CL}) and (\ref{eq:CLint}), for each magnitude using two different tetrahedral
samplings of 20$^3$ (dashed red) and 40$^3$ (solid black) in the 
triangulation of the Fermi surface.
 The HFSH modes are real and completely determined up to a sign factor. This is the reason why we show the absolute value of the coefficients.

The first component ($L$=1) of the modulus of the velocity (bottom) gives just the average value, as this mode is constant.
That fcc-Cu is a free-electron-like material is confirmed because the $L$=1 component is the strongest
contribution by far. 
Furthermore, 
the HFSH spectrum quantifies to which extent the modulus of the Fermi surface velocity is isotropic (or anisotropic). The strength of the peaks decreases 
rapidly and only some of the HFSH modes contribute significantly. We identify these modes graphically ($L$=1, 16, 34, 72 and 120) in the 
inset  of figure~\ref{fig:CuCL} (bottom).

The HFSH mode analysis for the $x$ component of the velocity (top) shows that the first component is equal to zero $c_{L=1}(v_x)=0$, 
while the next two modes $c_{L=2,3}(v_x)$ are the most important. This is consistent with the interpretation of the first three non-trivial 
degenerate modes to be similar to the ordinary $p$-like spherical harmonics 
(see figure~\ref{fig:CuPeriod}) and the $x$ direction (in our coordinate system) 
appears basically as a linear combination of the $L$=2 and $L$=3 HFSH modes. 
The average values over the Fermi surface of the $v_x$ and $v_x v_y$ functions are 
zero because these functions are obviously odd, and therefore 
 $c_{L=1}(v_x)=0$ and $c_{L=1}(v_x v_y)=0$, in both cases.

The $c_L$ components of the HFSH for the dense 40$^3$ (solid black) and coarser 20$^3$ 
(dashed red) samplings follow each other very closely. 
Of course, the HFSH energies are slightly reordered  going from a coarse to a denser mesh.
This is specially clear in the middle panel where we show the results of the $v_x v_y$ which has a more complicated spatial structure comparing to $v_x$ and $|v|$
and the amplitudes at higher energies are stronger. In any case, we find that at $L\sim 180$
 the  HFSH amplitudes ($c_L$)
are already about $20$ times weaker than the maximum values of $c_L$ at lower $L$. 

\begin{table}[h]
\caption{\label{tab:CLerrorsCu}%
{
Mismatch error (percentage) obtained using (\ref{eq:missm}) for $v_x$, $v_x  v_y$ and modulus of 
the velocity ($|v|$) in fcc-Cu as a function of the HFSH number of modes ($N_L$). 
}
}
\begin{indented}
\lineup
\item[]\begin{tabular}{c l l l l l l }
\br
  & \multicolumn{3}{c}{ \textrm{20$^3$}} &  \multicolumn{3}{c} {\textrm{40$^3$}}  \\
\cline{2-7}
 \textrm{$N_L$} &\multicolumn{1}{c}{$v_x$} &\multicolumn{1}{c}{$v_x v_y$} & \multicolumn{1}{c}{$|v|$} &\multicolumn{1}{c}{$v_x$} &\multicolumn{1}{c}{$v_x v_y$} & \multicolumn{1}{c}{$|v|$} \\
\mr
101 & 6.9 & 20.5 & 4.6    & 6.8 & 20.0 & 4.3    \\
201 & 3.1 & \06.9 & 2.1     & 2.7 & \05.5 & 1.8    \\
401 & 1.5  & \03.5 & 1.1    & 0.9   & \01.8 & 0.6        \\
701 & 1.2  & \02.6 & 0.5      & 0.4   & \00.9   & 0.3      \\
\br
\end{tabular}
\end{indented}
\end{table}

Table~\ref{tab:CLerrorsCu}  shows the analysis of the mode number dependence of the mismatch 
error for the $x$
cartesian component of the velocity ($v_x$), the product of the $x$ and $y$ components ($v_x v_y$) and the modulus of the velocity ($|v|$)
for the two different -tetrahedral- samplings considered in the marching tetrahedron method (20$^3$ and 40$^3$) for fcc-Cu.
We observe that the mismatch error is reduced by considering a denser triangular
 mesh, specially when a relatively large 
number of modes (see the rows corresponding to $N_L$=401 and 701  in table~\ref{tab:CLerrorsCu})
is used. The mismatch error for $v_x v_y$
is the largest because this function shows a richer spatial structure than $v_x$ or $|v|$ 
(see figure~\ref{fig:CuCL}). 
Expanding the functions with  $N_L\sim$701 modes is sufficient in order to obtain an estimated error below  $1\%$ with a tetrahedral 
sampling of 40$^3$. 

In fcc-Cu the number of vertices in the triangulation of the relatively coarse (20$^3$) and dense (40$^3$) tetrahedral 
meshes were  $N^{20^3}_v=6018$ and $N^{40^3}_v=24582$, respectively. 
One can estimate the data-storage saving when using the HFSH-mode representation
instead of the usual vertex ($\mathbf{k}$-space) representation. For example,
for a function such as  $F=v_x v_y$ and with a target error below $1\%$,
a saving factor of  the order of 
$\alpha({40^3})$$\sim$$1/40$ would be obtained with a 
 40$^3$ tetrahedral mesh.

\subsection{Magnesium diboride (hex-MgB$_2$)}

Hex-MgB$_2$ is an important example of strong electron-phonon coupling system \cite{Nagamatsu,Kortus,Bohnen} 
with a large superconducting transition temperature (for a BCS superconductor).

\begin{figure}
\centerline{\includegraphics[width=2.0\columnwidth]{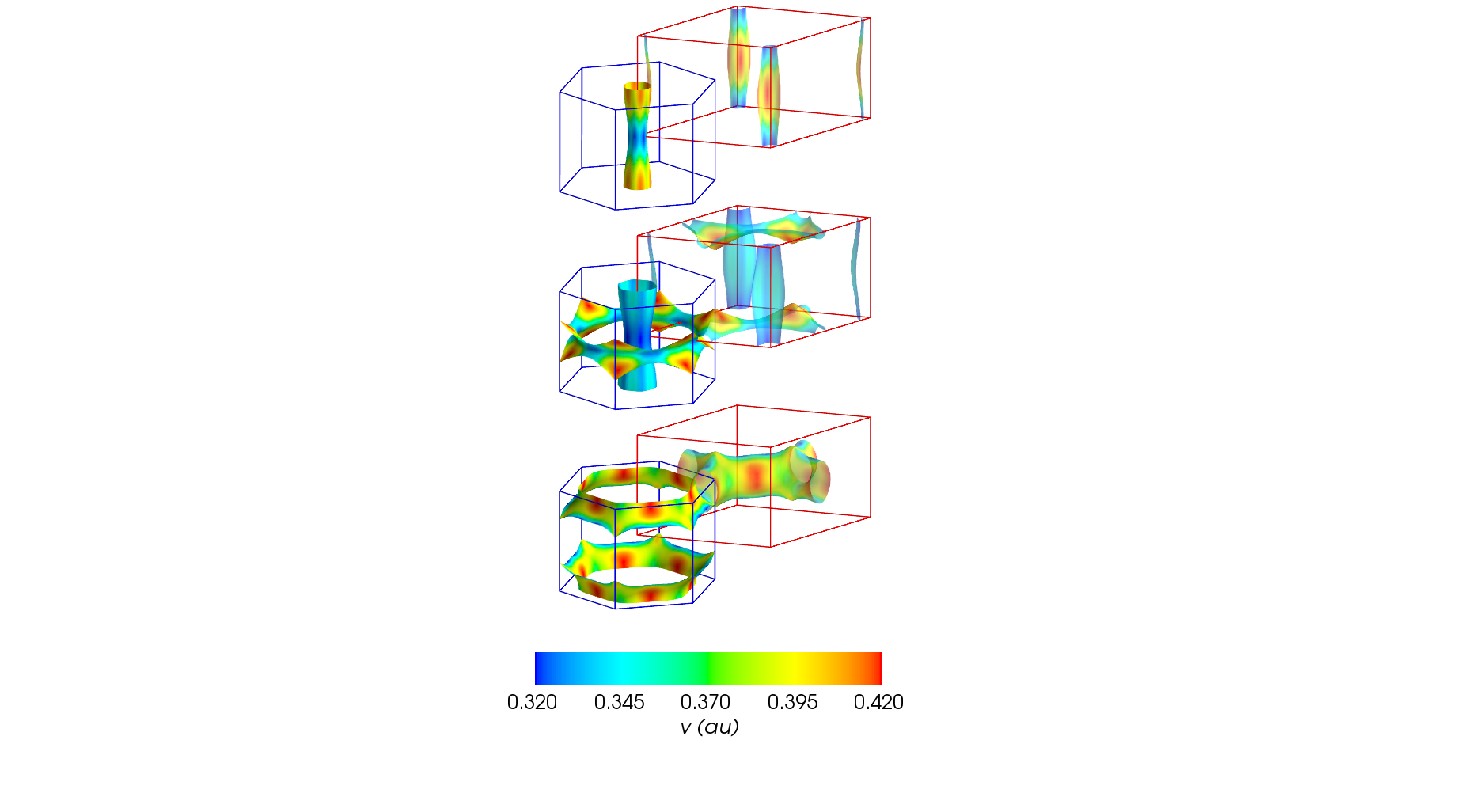}}
\caption{First (top), second (middle) and third (bottom) Fermi sheets  of magnesium diboride (hex-MgB$_2$) and the modulus of the electron velocity 
$v_{\mathbf{k}}$ (colour code) as a function of the electron momentum.
 The Fermi surface is shown for the
first Brillouin zone (blue) and the conventional zone (red),
 where it can be seen clearly why this surface presents genus $g$=2 for the
second and third bands.}
\label{fig:MgB2tri}
\end{figure}

This material presents three bands crossing the Fermi level,
 corresponding to the three Fermi sheets 
shown in figure~\ref{fig:MgB2tri}. 
All these surfaces are visualized for the
first Brillouin zone (blue) and the conventional cell or the polygon enclosed by the three reciprocal space lattice vectors
$\left (\mathbf{b}_1, \mathbf{b}_2, \mathbf{b}_3\right)$ (red).
The first band ($\sigma_1$) generates a single cylindroid Fermi surface (top). The
second band (middle) produces a Fermi surface which is in turn composed by two sheets:
 one of them ($\sigma_2$) produces
another cylindroid shape Fermi surface with a larger radius than that found in $\sigma_1$, 
and the second sheet ($\pi_1$) corresponds to a ring shape Fermi surface inside the first Brillouin zone.
The third band generates a single sheet ($\pi_2$) which appears as a double ring Fermi surface in the first
Brillouin zone, but translation to the conventional cell shows clearly 
that this surface is connected in a single sheet. 

From the point of view of the HFSH analysis hex-MgB$_2$ is a very interesting
test example  because
there are several features in this material that could potentially appear
 in more complex systems:
In hex-MgB$_2$ we have several bands (three) crossing the Fermi level, 
some of them  even composed by multiple sheets. Thus, this  system allows us
to demonstrate that the HFSH method is also applicable in a complex multiple 
Fermi sheet environment following exactly the same procedure as in simpler
 single-sheet Fermi surfaces, as those found in bcc-Li and fcc-Cu, for example.

The area,  the Euler characteristic ($\chi$) and genus ($g$) for
all the Fermi sheets of MgB$_2$ are shown in Table~\ref{tab:table1}.
 We emphasize again that in what concerns the DOS and the areas
of the different sheets of the Fermi surface, the coarser tetrahedral 
sampling (20$^3$) produces  practically converged results.
Moreover, we stress again that direct integration of (\ref{eq:DOS})
 (third column) compares very well with the linear 
tetrahedron method \cite{tetrahedrom}, as implemented in the Quantum
 Espresso code \cite{espresso}, which
makes us confident about the quality of the scalar products needed for the HFSH mode analysis 
(see sections \ref{propertiesHFSH}  and \ref{filtering}).

The $\sigma_1$ sheet of the Fermi surface (top of figure~\ref{fig:MgB2tri})
 is composed by a single sheet. Periodic boundary
conditions impose that any point in the surface $\mathbf{k}$ is equivalent to $\mathbf{k}+\mathbf{b}_3$, thus this surface sheet
is exactly a toroid from the topological point of view. 
This is confirmed
with the integer values of the Euler characteristic and genus obtained ($\chi=0$,
 $g=1$) up to numerical precision ($10^{-13}$) for this band.
The Euler characteristic is additive for surfaces which are not connected, as those
 found in the Fermi surface corresponding to the second 
band of MgB$_2$ (middle in figure~\ref{fig:MgB2tri}). In this surface we find that $\chi=0$ (genus $g=1$) for the central cylindroid  ($\sigma_2$) and
Euler characteristic $\chi=-2$ (genus $g=2$) for the outermost ring shape Fermi sheet 
described within the first Brillouin zone ($\pi_1$).
The third band of MgB$_2$ shows also a genus $g=2$ which is better understood 
when one looks at the Fermi surface plotted inside
 the conventional cell (red polyhedra in figure~\ref{fig:MgB2tri}). Periodic boundary conditions apply equally well for the conventional cell,
so that each neck (there are four) is connected with another by a simple lattice translation. Thus one concludes that the genus is equal to $g=2$.
A similar reasoning applies for the $\pi_1$ sheet of the second band.
Although topology is not our primary interest in this work, 
these results demonstrate the accuracy reached in the 
determination of the input data for the HFSH set in (\ref{eq:HE}) and 
(\ref{eq:HEb}).

\begin{figure}
\centerline{\includegraphics[width=0.8\columnwidth]{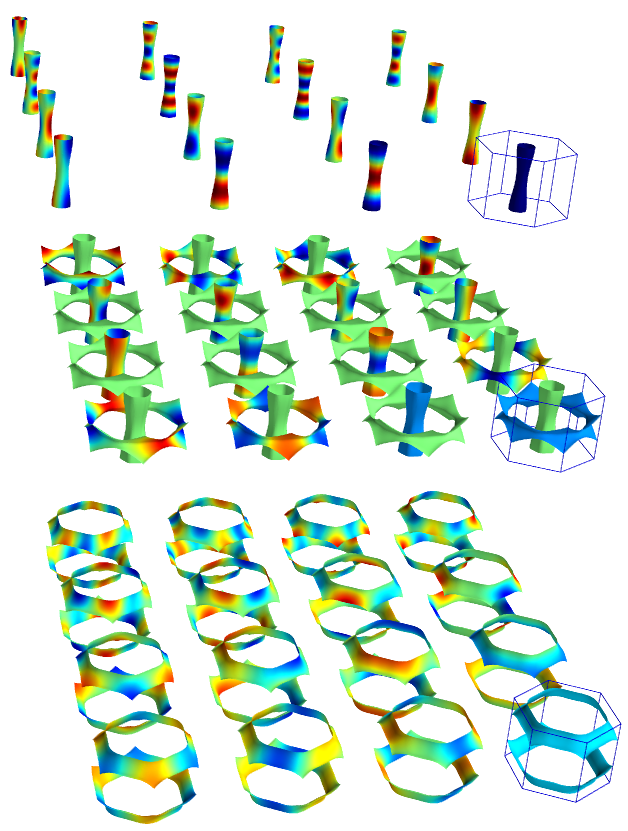}}
\caption{Lowest energy HFSH functions of for the three bands crossing the Fermi level in  hex-MgB$_2$.}
\label{fig:MgB2period}
\end{figure}

In figure~\ref{fig:MgB2period} we show the calculated HFSH set for the three  bands crossing the Fermi level for hex-MgB$_2$.
We show the 16 lowest energy states for each Fermi sheet: First band (top),
second band (middle) and third band (bottom).
For each band the mode with the lowest energy is that enclosed by the BZ. Then 
the energy of the modes increases from right to left within the rows. This 
way, the highest energy mode sits, for each band, at the upper left-hand corner.

The HFSH modes found for the first band (top) correspond basically to the 
solutions of the Helmholtz or stationary Shr\"odinger
equations in a torus except for the weighting factor introduced by the 
local electron velocity, the deformation and the curvature.
As the radius is smaller than the length of the cylindroid the first 
two non-trivial HFSH (second and third modes in the first row)
 are varying only in the axial direction while
the third  non-trivial HFSH mode (bottom left-hand corner)
is the first one with transversal variation.

As we have mentioned before, the two sheets  of the second band
 (middle of figure~\ref{fig:MgB2tri}) are obviously not connected,
 thus the discretized version of the Helmholtz equation, (\ref{eq:HE}), appears
as a block diagonal system. One strategy to solve the equation would be to
 separately diagonalize a version of (\ref{eq:HE}) for each of these 
two sheets. However, at this point it is more interesting to treat 
the system barely and ignoring the block diagonal structure
because in many complex systems we may find a situation where an obvious method 
for separating different sheets may not exist.
We observe that in this band (middle of figure~\ref{fig:MgB2period}), 
the lowest energy HFSH mode (inside the first BZ) corresponds to zero value for 
the $\pi_1$
Fermi sheet (out ring shape) and a finite constant value for 
the $\sigma_2$ sheet (cylindroid). The next HFSH mode in energy is also
trivial but it is  finite and constant for $\pi_1$ and equal to zero for $\sigma_2$.
 We would obtain the same result if we  separately
diagonalized each diagonal block of the generalized eigenvalue equations 
[(\ref{eq:HE}) for the HFSH or (\ref{eq:HEb}) for BHFSH]
 and we  energetically
ordered the summation of both subspaces. Thus, the present example for the second band 
enables us to demonstrate that 
the brute force diagonalization works equally well for systems in which a simple inspection is not enough for the separation of the Fermi surface
in different sheets. Or for those systems where the block diagonalization is not justified for the sake of computational simplicity. 

Table~\ref{tab:CLerrorsMgB2} shows a summary of the HFSH mode analysis for
 the modulus of the electron velocity ($|v|$) the
$x$ cartesian component of the velocity ($v_x$) and the multiplication of the $x$ and $y$ components of the velocity ($v_x v_y$). 
The results are presented in separate columns for each Fermi sheet.
 We conclude that about 600 HFSH modes are sufficient for 
a reasonable ($\epsilon$$\lesssim$$5\%$) representation of the  velocities and even of  the tensor $v_x v_y$.
Already with $\sim$2000 modes the method is able to capture the fine  details ($\epsilon$$\lesssim$$1\%$) of these testing example 
functions $|v|$, $v_x$ and $v_x v_y$ for all sheets. Going to each band in detail, we observe that in the first sheet 600 modes are more 
than sufficient even for a fine detailed description. A tetrahedral sampling of density 40$^3$ produced 3864 vertices in this sheet, so the 
saving factor for a target error of  ($\epsilon$$\lesssim$$1\%$) would be of about $\alpha_1(40^3) \simeq 1/13$ .
The Fermi surfaces corresponding to the second and third bands produced $N_v$=12558 and $N_v$=11478 vertices, respectively, 
in the triangulation process. Following a similar analysis, one obtains that the saving factor would be of the order of  $\alpha_2(40^3) \simeq 1/6$ and 
$\alpha_3(40^3) \simeq 1/9$ for the second and  third band, respectively.

\begin{table}
\caption{\label{tab:CLerrorsMgB2}%
{
Percentage of error obtained using  (\ref{eq:missm}) for $v_x$, $v_x  v_y$ and modulus of 
the velocity ($|v|$) as a function of the number of modes used in the summation for each of the Fermi crossing bands
in MgB$_2$. The dash means the error is lower than 0.1\%.
}
}
\begin{indented}
\lineup
\item[]\begin{tabular}{@{}*{10}{l} }
\br
  & \multicolumn{3}{c}{ \textrm{Band 1}} &  \multicolumn{3}{c}{ \textrm{Band 2}} & \multicolumn{3}{c}{ \textrm{Band 3}}  \\
\cline{2-10}
 \textrm{$N_L$} &\multicolumn{1}{c}{$v_x$} &\multicolumn{1}{c}{$v_x v_y$} & \multicolumn{1}{c}{$|v|$} &\multicolumn{1}{c}{$v_x$} &\multicolumn{1}{c}{$v_x v_y$} & \multicolumn{1}{c}{$|v|$} &\multicolumn{1}{c}{$v_x$} &\multicolumn{1}{c}{$v_x v_y$} &\multicolumn{1}{c}{$|v|$} \\
\mr
101 & 3.5 & 3.8 & 1.6   & 15.9 & 27.0 & 6.4     & 8.3 & 12.8 & 2.8 \\
601 & 0.2&0.5 & 0.1     & \01.9  & \05.0 & 0.6    & 1.5 & \02.1  & 0.4 \\
1201 & 0.1 &0.3 & ---   & \01.2  & \03.1 & 0.4    & 0.8 & \01.4  & 0.2 \\
1801 & --- &0.2 &  ---  & \00.9  & \02.1 & 0.3    & 0.2 & \00.5  & 0.1 \\
2001 & --- & --- & ---  & \00.3  & \01.2 & ---   & 0.1 & \00.3  &  ---  \\
\br
\end{tabular}
\end{indented}
\end{table}

\section{Conclusions}

We propose a new functional set, the Helmholtz Fermi Surface  Harmonics (HFSH), which shows very interesting properties
for efficiently representing physical quantities and/or integro-differential equations defined on the Fermi surface. 
This functional set is defined as the solutions of a Helmholtz type equation defined on top the Fermi surface, and we describe in detail 
the numerical scheme needed to solve this equation in a curved space including periodic boundary conditions. 
Although the HFSH presented in this work are defined very differently from the FSH proposed by
 Allen, all the analytical results for the Boltzmann equation and for the  anisotropic Eliashberg equation
reported in \cite{FSH-Allen,AllenMitrovic} are still  valid for  the HFSH presented in this work.

Despite the fact that topology is not the main goal of this work,  direct application of the Gauss-Bonnet theorem (integral 
of the Gaussian curvature) has
showed that the genus and the Euler characteristic of the Fermi surface
 is easily accessible only with the input data needed to set up the Helmholtz
 equation on the Fermi surface.
Thus, a systematic/automatic study of the topology of the Fermi surface of a material, for example as function of pressure, 
is accessible with this method  and without the need of any visual interpretation.

We have applied our method in bcc-Li, bcc-Na, fcc-Cu, fcc-Pb, bcc-W and hex-MgB$_2$, demonstrating that
a systematic procedure is easily implemented and that the method is robust, 
even in systems with complex band structures and/or several Fermi sheets.

We have expanded the cartesian components of the electron velocity and their
product in terms of HFSH, and we have found that  a relatively small
number ($\sim$10$^3$) of HFSH modes is enough for a very accurate description (error $\lesssim$1\%).
Thus, the HFSH mode representation enables to numerically tabulate any anisotropic magnitude defined on  the Fermi surface, 
allowing not only qualitative and faithful comparisons of these magnitudes calculated from different computational methods,
but also allowing the efficient integration of any anisotropic function over the Fermi surface.

%

\ack

The authors acknowledge the 
Department of Education, Universities and
Research of the Basque Government and UPV/EHU (Grant No. IT756-13) and the Spanish Ministry of Science 
and Innovation (Grant No. FIS2010-19609-C02-01) for financial support. Computer facilities were
provided by the Donostia International Physics Center (DIPC).

\appendix
\section{The marching tetrahedra algorithm}\label{marchingt}
\begin{figure}
\centerline{\includegraphics[width=0.5\columnwidth]{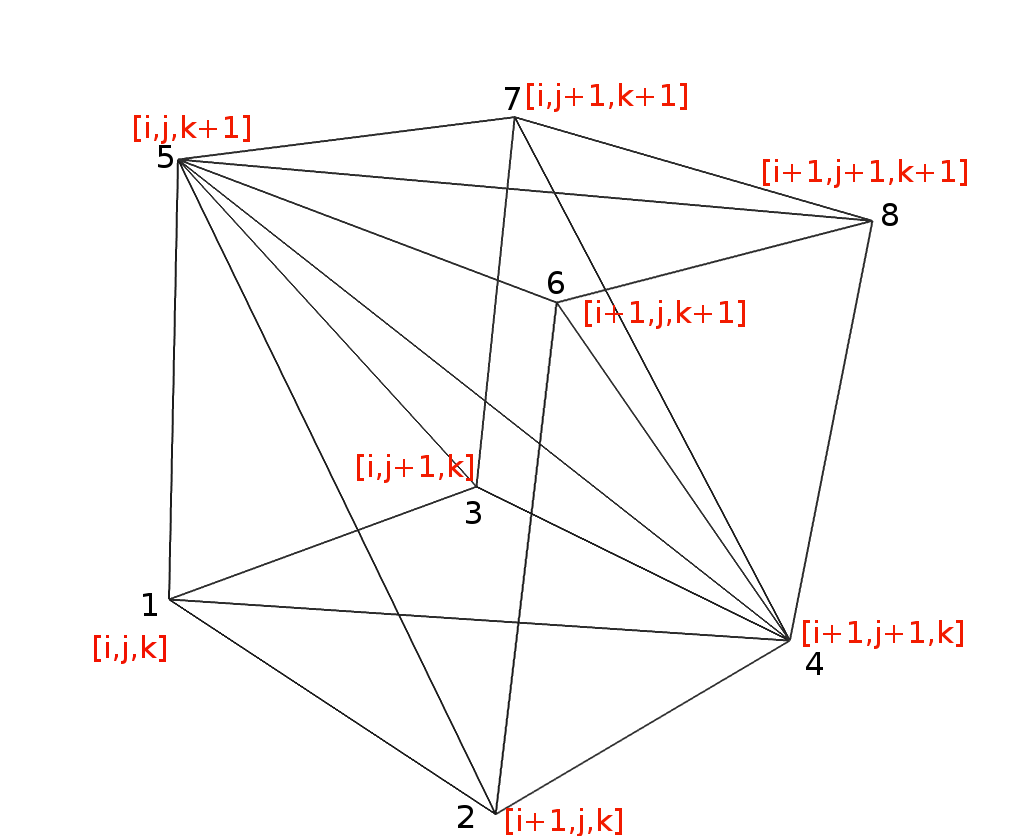}}
\caption{ 
Tetrahedral subdivision of a cube in terms of cube vertices. We consider 6 tetrahedra in each cube 
with labeling corner (i, j, k).) }
\label{fig:cube}
\end{figure}

The marching tetrahedra algorithm consists basically on dividing the entire Brillouin zone (BZ) in smaller
tetrahedral units such that the electron energy is linearly interpolated (inside each tetrahedra). 
Similar to the improved -or regular- tetrahedra method for integrations in the Brillouin zone \cite{tetrahedrom}, a regular
mesh of the first BZ is constructed as a first step and
the entire BZ is divided into $nk_1$$\times nk_2$$\times nk_3$ cubes of equal volume. Each 
of these cubes are labelled by the tag (i, j, k) -in crystal coordinates- closest to the $\Gamma$ point.
Subsequently, the volume of each cube is again subdivided in smaller tetrahedral simplexes. Among several possibilities,
we consider the 6 tetrahedra subdivision as illustrated in figure~\ref{fig:cube}. 

Let us for the moment suppose that the electron band energies are accessible in a 
$(nk_1+1)$$\times(nk_2+1)$$\times(nk_3+1)$ Monkhorst-Pack division of the BZ. All the $6 nk_1$$ nk_2$$nk_3$ tetrahedra
filling the BZ volume are constructed by means of the vertices located at the regular Monkhorst-Pack division (and eventually related by symmetry),
thus, the electron energies corresponding to all tetrahedral vertices are automatically known ($e_1$, $e_2$, $e_3$, $e_4$).

\begin{figure}[t]
\centerline{\includegraphics[width=0.5\columnwidth]{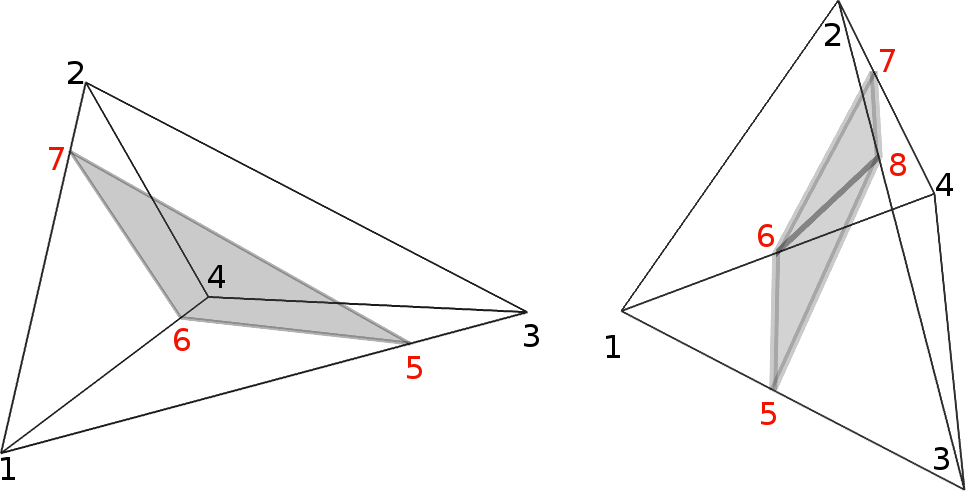}}
\caption{ 
Given the electron the list of electron energies calculated at the vertices of a tetrahedra $e_1$$\le$$e_2$$\le$$e_3$$\le$$e_4$ one finds
only two non-trivial intersection of the linearly interpolated energy inside the tetrahedra volume and the plane representing the Fermi energy $e_F$.
(Left) Single triangle in case of  $e_1$$\le$$e_F$$\le$$e_2$$\le$$e_3$$\le$$e_4$ or $e_1$$\ge$$e_F$$\ge$$e_2$$\ge$$e_3$$\ge$$e_4$. (Right) A rectangle
intersection represented by two triangles in the case of  $e_1$$\le$$e_2$$\le$$e_F$$\le$$e_3$$\le$$e_4$.}
\label{fig:tetra}
\end{figure}

The triangulation of the Fermi surface is calculated by checking, in a first step, that the Fermi level ($e_F$)
is located between the minimum and the maximum of the energies corresponding to all the tetrahedral vertices. That is, 
if $\mbox{Min}(e_i)<e_F<\mbox{Max}(e_i), i=1,4$ part of the Fermi surface is found in the interior of a given tetrahedra.
Let us suppose for simplicity, that the energies are ordered in increasing order
$e_1$$\le$$e_2$$\le$$e_3$$\le$$e_4$. 
In this case, one can only  find two nontrivial possibilities, 
(i) corresponding to the simple triangular section when
 $e_1$$\le$$e_F$$\le$$e_2$$\le$$e_3$$\le$$e_4$
or $e_1$$\le$$e_2$$\le$$e_3$$\le$$e_F$$\le$$e_4$ (left side of figure~\ref{fig:tetra}),
 and (ii) the case
when the intersection is quadrilateral $e_1$$\le$$e_2$$\le$$e_F$$\le$$e_3$$\le$$e_4$ (right side figure~\ref{fig:tetra}),
and the two triangle simplexes are generated. There are two possibilities to divide a quadrilateral by triangles, among
them, our choice is the one in which both triangles have an area as similar as possible. 

In all cases the vertices of the  intersection are easily calculated by considering that the electron energy 
is a linear function inside each tetrahedra. The vertices of the -Fermi surface- intersection triangle 
corresponding to the case 1 ( $e_1$$\le$$e_F$$\le$$e_2$$\le$$e_3$$\le$$e_4$) are given by the vectors  
\begin{eqnarray}
\nonumber
\mathbf{t_5} &=& \mathbf{t_1}+\frac{e_F-e_1}{e_3-e_1}  \left( \mathbf{t_3} - \mathbf{t_1} \right ) \ \ \ \ \ \ \ \ \ \ \ \\
\nonumber
\mathbf{t_6} &=& \mathbf{t_1}+\frac{e_F-e_1}{e_4-e_1}  \left( \mathbf{t_4} - \mathbf{t_1} \right ) \\
\nonumber
\mathbf{t_7} &=& \mathbf{t_1}+\frac{e_F-e_1}{e_2-e_1}  \left( \mathbf{t_2} - \mathbf{t_1} \right ).
\end{eqnarray}

The case $e_1$$\le$$e_2$$\le$$e_3$$\le$$e_F$$\le$$e_4$ is equivalent to the one above by considering the opposite -descending- energy 
ordering. The vectors describing the quadrilateral intersection in case 2 ($e_1$$\le$$e_2$$\le$$e_F$$\le$$e_3$$\le$$e_4$) are  similarly obtained,
\begin{eqnarray}
\nonumber
\mathbf{t_5} &=& \mathbf{t_1}+\frac{e_F-e_1}{e_3-e_1}  \left( \mathbf{t_3} - \mathbf{t_1} \right ) \ \ \ \ \ \ \ \ \ \ \ \\
\nonumber
\mathbf{t_6} &=& \mathbf{t_1}+\frac{e_F-e_1}{e_4-e_1}  \left( \mathbf{t_4} - \mathbf{t_1} \right ) \\
\nonumber
\mathbf{t_7} &=& \mathbf{t_2}+\frac{e_F-e_1}{e_4-e_2}  \left( \mathbf{t_4} - \mathbf{t_2} \right ) \\
\nonumber
\mathbf{t_8} &=& \mathbf{t_2}+\frac{e_F-e_1}{e_3-e_2}  \left( \mathbf{t_3} - \mathbf{t_2} \right ),
\end{eqnarray}
and the triangle vertices would correspond to (5,8,6) and (6,8,7), or alternatively to (5,7,6) and (5,8,7). 
Our choice is the one in which both areas a maximally similar.

\section{The maximally localized Wannier approach for calculating electron energies and velocities}\label{sec:wannier}

The Wannier method is used extensively in the present work as an efficient input tool for calculating the electron energy 
and velocities. In this section we introduce this method only succinctly for completeness,
 and refer the reader to the original works \cite{wannier-original,wannier-entangled,wannier90}
for more details.

In the Wannier scheme as considered in this article, the first step is to perform an ordinary
 DFT electronic structure calculation on a regular and relatively coarse Monkhorst-Pack $\mathbf{k}$ mesh 
of density $n_1 \times n_2 \times n_3$ in order to obtain
the energies $\epsilon_{\mathbf{k},i}$ and eigenfunctions $\Phi_{\mathbf{k},i}(\mathbf{r})$ for the given BZ division
and band index $i$. 

The maximaly localized Wannier functions are  calculated (and defined) by considering a unitary 
transformation (mixing) among Bloch states and minimizing a spread functional. 
Considering, for simplicity, $n$ isolated bands in each $\mathbf{k}$ point one considers a general lattice periodic function as
\begin{eqnarray}\label{eq:blochW}
\psi_{\mathbf{k},\alpha}(\mathbf{r}) \equiv \sum_{\beta} U_{\mathbf{k}, \alpha, \beta} \Phi_{\mathbf{k},\beta}(\mathbf{r}),
\end{eqnarray}
where $U_{\mathbf{k}, \alpha, \beta}$ represents -in principle- any unitary matrix of dimension $n$. In this method, however,
one is interested in making the functions $\psi_{\mathbf{k},\alpha}$ as maximally flat as possible, such that 
any interpolation procedure works out effectively. Since maximally flat in reciprocal space is equivalent to maximally localized in real 
space, the $U_{\alpha, \beta}$ unitary matrices are fixed by minimizing the spread functional
\begin{eqnarray}\label{eq:spread}
\Omega \equiv \sum_{\alpha} \left (  \langle \chi_{\alpha} | r^2 | \chi_{\alpha}  \rangle - \langle \chi_{\alpha}  | r | \chi_{\alpha} \rangle^2 \right),
\end{eqnarray}
where the Wannier functions are defined by  
\begin{eqnarray}\label{eq:W}
\chi_{\alpha} (\mathbf{r-R}) \equiv \frac{1}{N_k} \sum_{\mathbf{k}} \psi_{\mathbf{k},\alpha} (\mathbf{r}) e^{i \mathbf{k} \mathbf{R} },
\end{eqnarray}
with the unitary matrices in (\ref{eq:blochW})  fixed by minimizing the spread $\Omega$ in (\ref{eq:spread}).
In the equation above the summation in $\mathbf{k}$ goes over the initial 
 Monkhorst-Pack coarse mesh  of density $n_1 \times n_2 \times n_3$ considered for the self consistent
DFT calculation.

\subsection{Fourier interpolation of the Wannier hamiltonian}

As mentioned before, the Bloch functions in (\ref{eq:blochW}) are maximally flat in reciprocal space -by construction in the 
Wannier procedure-, thus the Fourier interpolation scheme for any quantity involving the electron wave functions is very effective.
The electron band energies and velocities are very efficiently calculated by considering the electronic DFT hamiltonian in the basis
of Wannier Bloch functions $\psi_{\mathbf{k},\alpha}$

\begin{eqnarray}\label{eq:HW}
H^W_{\mathbf{k}, \alpha,\beta} %
=\sum_{\mathbf{R}} e^{i \mathbf{k} \mathbf{R}} \langle \chi_{\alpha} (\mathbf{r}) | H_{DFT} |   \chi_{\beta} (\mathbf{r-R}) \rangle.
\end{eqnarray}
Note that in (\ref{eq:HW}) $\mathbf{k}$ is not restricted to the self consistent DFT coarse mesh and any 
$\mathbf{k}$ vector is accessible, making the Fourier interpolation procedure straightforward. Considering any  
$\mathbf{k}$ wave vector in reciprocal space $H^W_{\mathbf{k}, \alpha,\beta}$ is accessible and the Fourier
interpolated energies ($\epsilon_{\mathbf{k}, i}$) are obtained by an ordinary diagonalization procedure,
\begin{eqnarray}\label{eq:UHWU}
\sum_{\alpha,\beta} U^{-1}_{\mathbf{k}, i,\alpha} H^W_{\mathbf{k}, \alpha,\beta}  U_{\mathbf{k}, \beta, j} = \delta_{i,j} \epsilon_{\mathbf{k}, i}.
\end{eqnarray}

The velocities are similarly calculated by considering the expectation value of the gradient of the Wannier hamiltonian and the solution 
eigenvectors in (\ref{eq:UHWU}),
\begin{eqnarray}
\vec{v}_{\mathbf{k}, i} =  \sum_{\alpha,\beta} U^{-1}_{\mathbf{k}, i, \alpha} \left ( 
\sum_{\mathbf{R}}  i \vec{R} \  e^{i \mathbf{k} \mathbf{R}} 
\langle \chi_{\alpha, \mathbf{0}}  | H_{DFT} | \chi_{\beta, \mathbf{R}}  \rangle \right ) U_{\mathbf{k}, i, \beta}. \ \ \ \ \ 
\end{eqnarray}

\section*{References}
\bibliographystyle{unsrt}
\bibliography{biblio}

\end{document}